\documentclass[times, twocolumn]{aastex631}
\usepackage{graphicx, graphics}
\usepackage{CJK}
\usepackage{bbm}
\usepackage{comment}
\definecolor{cyan(process)}{rgb}{0.0, 0.72, 0.92}

\usepackage[T1]{fontenc} % enable polish characters for Lau 2022 citation

\usepackage{amsmath, amssymb, bm}
\usepackage[nodayofweek]{datetime}
\usepackage{multirow}
\input{prepcitationformat.txt}

\makeatletter
\newcommand*{\centerfloat}{%
  \parindent \z@
  \leftskip \z@ \@plus 1fil \@minus \textwidth
  \rightskip\leftskip
  \parfillskip \z@skip}
\makeatother

\newdateformat{monthyearday}{%
  \THEYEAR\ \monthname[\THEMONTH] \twodigit{\THEDAY}}

\newcommand{\uat}[2]{\href{http://vocabs.ands.org.au/repository/api/lda/aas/the-unified-astronomy-thesaurus/current/resource.html?uri=http://astrothesaurus.org/uat/#1}{#2 (#1)}}

\received{2023 January 31}
\revised{\monthyearday\today}
\accepted{}
\published{}
\submitjournal{AAS Journals}

\newcommand{\obj}{2MASS J04124068+2438157}
\newcommand{\shortnameobj}{2M041240}
\newcommand{\affilCaltechAstro}{\affiliation{Department of Astronomy, California Institute of Technology, MC 249-17, 1200 East California Boulevard, Pasadena, CA 91125, USA}}

\shorttitle{A Double-Ring Disk}
\shortauthors{Long et al.}

\graphicspath{{./}{figures/}}

\begin{document}
\begin{CJK*}{UTF8}{gbsn}
%\title{A Large Double-ring Disk around a Low Mass Star}
%\title{A Large Double-ring Disk around a mid-M Dwarf 2MASS J04124068+2438157}
\title{A Large Double-ring Disk around the Taurus M Dwarf J04124068+2438157}

\author[0000-0002-7607-719X]{Feng Long (龙凤)}
\altaffiliation{NASA Hubble Fellowship Program Sagan Fellow}
\affiliation{Lunar and Planetary Laboratory, University of Arizona, Tucson, AZ 85721, USA; \url{fenglong@arizona.edu}}
\affiliation{Center for Astrophysics \textbar\, Harvard \& Smithsonian, 60 Garden St., Cambridge, MA 02138, USA}

\author[0000-0003-1698-9696]{Bin B. Ren (任彬)} % NIRC2 observation and reduction 
\affiliation{Universit\'{e} C\^{o}te d'Azur, Observatoire de la C\^{o}te d'Azur, CNRS, Laboratoire Lagrange, F-06304 Nice, France}
\affiliation{Universit\'{e} Grenoble Alpes, CNRS, Institut de Plan\'{e}tologie et d'Astrophysique (IPAG), F-38000 Grenoble, France}
\affilCaltechAstro

\author[0000-0003-0354-0187]{Nicole L. Wallack} % NIRC2 contrast curve and mass limits
\affiliation{Earth and Planets Laboratory, Carnegie Institution for Science, Washington, DC 20015, USA}
\affiliation{Division of Geological \& Planetary Sciences, California Institute of Technology, MC 150-21, 1200 East California Boulevard, Pasadena, CA 91125, USA}

\author[0000-0001-6307-4195]{Daniel Harsono}
\affiliation{Institute of Astronomy, Department of Physics, National Tsing Hua University, Hsinchu, Taiwan}

\author[0000-0002-7154-6065]{Gregory J. Herczeg (沈雷歌)}
\affiliation{Kavli Institute for Astronomy and Astrophysics, Peking University, Beijing 100871, China}

\author{Paola Pinilla}
\affiliation{Mullard Space Science Laboratory, University College London, Holmbury St Mary, Dorking, Surrey RH5 6NT, UK}

\author[0000-0002-8895-4735]{Dimitri Mawet} % NIRC2 proposal PI at Caltech
\affilCaltechAstro
\affiliation{Jet Propulsion Laboratory, California Institute of Technology, 4800 Oak Grove Drive, Pasadena, CA 91109, USA}

\author[0000-0003-2232-7664]{Michael C. Liu} % NIRC2 proposal PI at U Hawaii, actual PI for the night of this target
\affiliation{Institute for Astronomy, University of Hawaii, 2680 Woodlawn Drive, Honolulu, HI 96822, USA}

% The rest of Keck team can go alphabetical, names and affiliations will be provided
\author[0000-0003-2253-2270]{Sean M. Andrews}
\affiliation{Center for Astrophysics \textbar\, Harvard \& Smithsonian, 60 Garden St., Cambridge, MA 02138, USA}

\author[0000-0001-6906-9549]{Xue-Ning Bai}
\affiliation{Institute for Advanced Study and Department of Astronomy, Tsinghua University, 100084, Beijing, China}

\author{Sylvie Cabrit}
\affiliation{Sorbonne Universit\'{e}, Observatoire de Paris, Universit\'{e} PSL, CNRS, LERMA, F-75014 Paris, France}
\affiliation{Universit\'{e} Grenoble Alpes, CNRS, Institut de Plan\'{e}tologie et d'Astrophysique (IPAG), F-38000 Grenoble, France}

\author[0000-0002-2828-1153]{Lucas A. Cieza}
\affiliation{Instituto de Estudios Astrof\'isicos, Facultad de Ingenier\'ia y Ciencias, Universidad Diego Portales, Av. Ejercito 441, Santiago, Chile}
\affiliation{Millennium Nucleus on Young Exoplanets and their Moons (YEMS)}

\author[0000-0002-6773-459X]{Doug Johnstone}
\affiliation{NRC Herzberg Astronomy and Astrophysics, 5071 West Saanich Rd, Victoria, BC, V9E 2E7, Canada}
\affiliation{Department of Physics and Astronomy, University of Victoria, Victoria, BC, V8P 5C2, Canada}

\author[0000-0002-0834-6140]{Jarron M. Leisenring}
\affiliation{Steward Observatory, University of Arizona, 933 N Cherry Ave, Tucson, AZ 85721, USA}

\author{Giuseppe Lodato}
\affiliation{Dipartimento di Fisica, Universita Degli Studi di Milano, Via Celoria, 16, I-20133 Milano, Italy}

\author{Yao Liu}
\affiliation{Purple Mountain Observatory \& Key Laboratory for Radio Astronomy, Chinese Academy of Sciences, Nanjing 210023, China}

\author[0000-0003-3562-262X]{Carlo F. Manara}
\affiliation{European Southern Observatory, Karl-Schwarzschild-Str. 2, D-85748 Garching bei M\"{u}nchen, Germany}

\author[0000-0002-1078-9493]{Gijs D. Mulders}
\affiliation{Facultad de Ingenier\'ia y Ciencias, Universidad Adolfo Ib\'a\~nez, Av.\ Diagonal las Torres 2640, Pe\~nalol\'en, Santiago, Chile}
\affiliation{Millennium Institute for Astrophysics, Chile}

\author[0000-0001-5378-7749]{Enrico Ragusa}
\affiliation{Univ Lyon, Univ Lyon1, Ens de Lyon, CNRS, Centre de Recherche Astrophysique de Lyon UMR5574, F-69230 Saint-Genis-Laval, France}

\author{Steph Sallum}
\affiliation{Department of Physics and Astronomy, University of California, Irvine, 4129 Frederick Reines Hall, Irvine, CA 92697-4575, USA}

\author[0000-0001-9277-6495]{Yangfan Shi}
\affiliation{Kavli Institute for Astronomy and Astrophysics, Peking University, Beijing 100871, China}

\author[0000-0003-3590-5814]{Marco Tazzari}
\affiliation{Institute of Astronomy, University of Cambridge, Madingley Road, CB3 0HA Cambridge, UK}

\author[0000-0002-6879-3030]{Taichi Uyama}
\affiliation{Infrared Processing and Analysis Center, California Institute of Technology, 1200 East California Boulevard, Pasadena, CA 91125, USA}
\affiliation{NASA Exoplanet Science Institute, Pasadena, CA 91125, USA}
\affiliation{National Astronomical Observatory of Japan, 2-21-1 Osawa, Mitaka, Tokyo 181-8588, Japan}

\author[0000-0002-4309-6343]{Kevin Wagner}
\altaffiliation{NASA Hubble Fellowship Program Sagan Fellow}
\affiliation{Steward Observatory, University of Arizona, 933 N Cherry Ave, Tucson, AZ 85721, USA}

\author[0000-0003-1526-7587]{David J. Wilner}
\affiliation{Center for Astrophysics \textbar\, Harvard \& Smithsonian, 60 Garden St., Cambridge, MA 02138, USA}

\author[0000-0002-6618-1137]{Jerry W. Xuan}
%\affiliation{Department of Astronomy, California Institute of Technology, Pasadena, CA 91125, USA}
\affilCaltechAstro

\begin{abstract}
Planet formation imprints signatures on the physical structures of disks. In this paper, we present high-resolution ($\sim$50\,mas, 8\,au) Atacama Large Millimeter/submillimeter Array (ALMA) observations of 1.3\,mm dust continuum and CO line emission toward the disk around the M3.5 star \obj. The dust disk consists only of two narrow rings at radial distances of 0$\farcs$47 and 0$\farcs$78 ($\sim$70 and 116\,au), with Gaussian $\sigma$ widths of 5.6 and 8.5\,au, respectively. 
The width of the outer ring is smaller than the estimated pressure scale height by $\sim25\%$, suggesting dust trapping in a radial pressure bump. The dust disk size, set by the location of the outermost ring, is significantly larger (by $3\sigma$) than other disks with similar millimeter luminosity, which can be explained by an early formation of local pressure bump to stop radial drift of millimeter dust grains.
After considering the disk's physical structure and accretion properties, we prefer planet--disk interaction over dead zone or photoevaporation models to explain the observed dust disk morphology.
We carry out high-contrast imaging at $L'$ band using Keck/NIRC2 to search for potential young planets, but do not identify any source above $5\sigma$. Within the dust gap between the two rings, we reach a contrast level of $\sim$7\,mag, constraining the possible planet below $\sim$2--4\,$M_{\rm Jup}$. Analyses of the gap/ring properties suggest a $\sim$Saturn mass planet at $\sim$90\,au is likely responsible for the formation of the outer ring, which can be potentially revealed with JWST.

\end{abstract}

\section{Introduction} \label{sec:intro}
Our observational understanding of planet formation in circumstellar disks has greatly advanced in the past decade, largely due to the advent of the Atacama Large Millimeter/submillimeter Array (ALMA). On one hand, near-complete ALMA disk surveys in a number of star-forming regions with moderate spatial resolution have revealed important trends among the global stellar and disk properties, for example, suggesting higher mass stars host more massive disks thus with larger potential to form giant planets (e.g., \citealt{Ansdell2016, Pascucci2016, RuzRodrguez2018, Cazzoletti2019}, see also pre-ALMA results in \citealt{Andrews2013}). On the other hand, high resolution imaging towards a targeted group of disks has shown that large and massive disks are often associated with substructures, mostly seen as gaps and rings (e.g., \citealt{ALMAPartnership2015, Andrews2018, Long2018, Cieza2021}), but also as arcs and spiral arms (\citealt{vanderMarel2013Sci...340.1199V, Dong2018_MWC758, Huang2018_spirals}). Given the ubiquitous nature of planetary systems in the Galaxy (e.g., \citealt{WinnFabrycky2015}), the frequent appearance of disk substructures suggests that they might be relevant to the process of planet formation, though establishing the direct connection between a disk feature and a planet is so far challenging, considering the complex physics involved in disk and planet evolution (e.g., \citealt{Nayakshin2020}).

These substructures provide an immediate observational solution to the long-standing problem in dust evolution and planet formation -- the so-called radial drift barrier \citep{Weidenschilling1977, Nakagawa1986}. In the default assumption of a smooth gas disk, millimeter-sized particles tend to move towards the (global) pressure maximum (the inner disk region) due to aerodynamic drag and quickly deplete the outer disk, in contradiction to early observations of large millimeter disks. Those recently identified dust rings/gaps, as regions of material accumulation/depletion, suggest the presence of local pressure maxima that stop the inward migration of particles and sustain them locally \citep{Pinilla2012_pressurebump}. For this reason, these are also favorable places of planet formation through planetesimal and/or pebble accretion (e.g., \citealt{Cummins2022, Jiang2022}).

Pressure bumps can result from planet--disk interactions, where the planet motion carves the disk material and builds some over-dense regions outside its orbit (e.g., \citealt{Rice2006, Dodson-Robinson2011, Pinilla2012_planet, Zhu2012, Paardekooper2022}). Many other mechanisms, including a variety of (magneto-) hydrodynamic instabilities, have also been proposed to produce pressure modulations and then trap particles (see references in \citealt{Andrews2020, Bae2022}). Though in most cases the origin of disk substructures is still debated, they are likely fundamental to the planet formation process. 
Thus, characterizing disk substructures in different star and disk environments is crucial in building a complete view of planet formation.

Currently, most high-resolution observations are designed to target disks around early type stars (M3 or earlier) and/or bright disks (that also mostly surround solar-mass stars, see references above). Though gaps and rings have been reported in a few bright M-dwarfs disks (\citealt{Kurtovic2021, Hashimoto2021, Pinilla2022EPJP}), our generic knowledge about dust substructure properties in disks around these late type stars is still missing. In addition, we expect that the dust radial drift problem is even more severe for disks around lower mass stars, as the drift velocity is faster when surrounding lower mass stars ($v_{\rm drift}\propto L_{*}^{1/4}/\sqrt{M_{*}}$, which is mostly controlled by stellar mass in the low mass regime, \citealt{Pinilla2013}). To understand how M-dwarf disks overcome the radial drift barrier and how substructure properties vary across stellar mass, we have conducted a high-resolution ALMA survey towards a number of M-dwarf disks in the Taurus star-forming region (Shi et al.~in prep). 

This paper reports on the interesting object \obj\ in our M-dwarf disk sample, which hosts a very extended dust disk composed of two dust rings. We arrange the paper as follows. Section~\ref{sec:source} provides a detailed summary of the source properties. Section~\ref{sec:obs} describes the ALMA and Keck observations, as well as their corresponding data reduction. Section~\ref{sec:result} presents our characterization of the disk morphology, and Section~\ref{sec:diss} discusses this target in the context of other disks and the origin of the dust rings. A summary is then given in Section~\ref{sec:sum}.

\section{Source Properties} \label{sec:source}

\obj\ (hereafter \shortnameobj) was first identified as a member of Taurus in a search for objects with mid-IR excess emission in the WISE survey (\citealt{rebull11}, as Class II disks).  Membership is confirmed with Gaia DR2 proper motions \citep{galli19}.  The Gaia EDR3 parallax leads to a distance of $148.7\pm0.5$ pc \citep{gaiadr3}.

We adopt the spectral type of M3.5\footnote{The stellar parameters adopted for this paper were based on measurements from the literature.  In Shi et al.~(in preparation), we are re-evaluating the stellar properties with a new spectral analysis and measure a spectral type of M4.3, which if adopted here would decrease the 50\% spotted mass to 0.30 M$_\odot$.  The main results from this paper are unchanged.} and the extinction $A_J=0.37$ from the compilation of Taurus members by \citet{esplin19}. 
The spectral type corresponds to an effective temperature of ${\sim}3300$ K \citep{herczeg14}, consistent with temperatures measured from LAMOST spectra \citep{luo22}.  The luminosity of \shortnameobj\ is $0.153$ $L_\odot$, measured from the 2MASS $J$-band magnitude of 11.151, the bolometric correction from \citet{pecaut13}, and zero-point flux of $3.013\times10^{35}$ erg/s.
The temperature and luminosity lead to a radius of 1.20 $R_\odot$ and correspond to a mass of $0.25~M_\odot$ and an age of 1.6 Myr using the \citet{baraffe15} and spotless \citet{somers20} models. 
When relying on \citet{feiden16} magnetic models of pre-main sequence evolution and the \citet{somers20} models for a young star with 50\% spot coverage, we obtain a mass of $0.38~M_\odot$ and an age of 4.4 Myr, and $0.42$ $M_\odot$ and 4.7 Myr, respectively. 
The ${\sim}0.4~M_\odot$ estimate is adopted here, given the consistency of those evolutionary tracks with the dynamical measurement of stellar mass for two M4 members of Taurus \citep{pegues21}. 
The sizes of dust disks around most stars are well within 100\,au, which also gradually decrease with stellar mass \citep{Hendler2020}, thus the finding of a $\sim120$\,au disk (see results in Section~\ref{sec:dust}) around this mid-M star is rather surprising and studied in detail here.

\shortnameobj\ is actively accreting, as indicated from the H$\alpha$ equivalent widths of 42 and 46 \AA\, measured by LAMOST on two different nights \citep{luo22}.  The \ion{He}{1} $\lambda6678$ equivalent width is 0.64 \AA\ in a Keck/HIRES high-resolution optical spectrum obtained on UT 2019 November 29 (PI: L.~Hillenbrand).  This equivalent width translates to a flux of  ${\sim}1.3\times10^{-15}$ erg cm$^{-2}$ s$^{-1}$, calculated based on the continuum flux obtained by flux-calibrating a low-resolution optical spectrum obtained from UH88/SNIFS \citep[see description in][]{guo18} with Gaia DR3 spectrophotometry. This flux should be robust to continuum variability, since the Gaia $R_P$-band flux varies by ${\sim}0.04$ mag \citep[ignoring one outlier,][]{gaiadr3}.  After correcting for extinction and distance, the line luminosity is $2.4\times 10^{-6}$ $L_\odot$, which corresponds to an accretion luminosity of 0.012 $L_\odot$ \citep[from the correlations of][]{alcala17} and a mass accretion rate of ${\sim}1.4\times10^{-9}~M_\odot$ yr$^{-1}$.  This accretion rate is likely variable, as the H$\alpha$ equivalent width in the epoch of Keck observation is 30 \AA, lower than those measured in the two epochs with LAMOST. 
The [\ion{O}{1}] $\lambda6300$ line has an equivalent width of 1.34 \AA\ and a FWHM of 32 km s$^{-1}$, indicating the presence of a disk wind \citep[e.g.][]{banzatti19}.

%%%%%%%%%%%%%%%%%%%%%%%%%%%%%%%%%
\section{Observations} \label{sec:obs}

\subsection{ALMA Observations}
\shortnameobj\ was observed with ALMA Band~6 receivers on UT 2021 August 27,  during the Return to Operations phase after the Covid19 shutdown, as part of program 2019.1.00566.S (PI: G.~Herczeg). The observations were performed with 39 antennas spanning baselines from 92\,m to 10.8\,km, with a total on-source time of 16.8\,min. The receivers were configured into four spectral windows (SPWs), including two continuum SPWs centered at 217 and 234.4\,GHz, each with a bandwidth of 1.875\,GHz. 
The two remaining SPWs targeted $^{12}$CO, $^{13}$CO, and C$^{18}$O $J=2-1$ lines at a channel spacing of 0.244\,MHz ($\sim$0.3\,km\,s$^{-1}$).

The raw visibilities were first pipeline calibrated using the specified CASA version 6.1.1 \citep{CASATeam2022PASP} to flag problematic data segments, correct for bandpass responses, set absolute flux scales, and solve for complex gain variations. We inspected the calibrated visibilities and identified some residual features of atmospheric absorption correction around channel 500 in the 234.4\,GHz SPW. The corresponding channel range of 400--600 was then flagged. 
Finally, one round of phase-only self-calibration (\texttt{solint=`inf'}) was performed. As the improvement of continuum image quality is subtle (${\sim}10\%$ in peak S/N), self-calibration solutions were not applied to the line SPWs.

Continuum images at a mean frequency of 225.6\,GHz (1.3\,mm) were generated with the \texttt{tclean} task.
To better visualize the faint disk millimeter emission, we produced two images with synthesized beam sizes of $0\farcs06\times0\farcs05\,(\rm PA=17 \degr)$ (\texttt{robust=0.5}) and $0\farcs12\times0\farcs12\,(\rm PA=26 \degr)$ (\texttt{robust=2}), where $uv$ taper is applied for a more circular beam\footnote{The Gaussian $uv$ tapers are $0\farcs04\times0\farcs0\,(\rm PA=-50 \degr)$ and $0\farcs08\times0\farcs04\,(\rm PA=-65 \degr)$ for images with 60 and 120\,mas resolution.}. The 1$\sigma$ noise levels computed from nearby emission-free regions in the images are $\sim$32 and 35\,$\mu$Jy\,beam$^{-1}$, respectively. The three CO lines were imaged at a velocity resolution of 0.4\,km\,s$^{-1}$ with a coarser beam size of $0\farcs2$, reaching a 1$\sigma$ noise level of $\sim$5-6\,mJy\,beam$^{-1}$ in individual channels.

\subsection{Keck/NIRC2 High-contrast Imaging Observations}
We used the Keck/NIRC2 vortex coronagraph \citep{mawet17, serabyn17} to suppress central starlight and observed the surroundings of \shortnameobj\ in $L'$-band (${\sim}3.78~\mu$m) on UT 2021 October 27 
%from 12:01:22 to 13:55:41 
under Keck program H290 (PI: M.~Liu). 
The \textit{Gaia}~DR3 $R_P$-band magnitude of 13.27 for \shortnameobj\ \citep{gaiadr3} is too faint for existing Shack-Hartmann wavefront sensors that perform adaptive optics corrections in visible wavelengths to efficiently conduct high-contrast imaging for companions. In comparison, its $H$-band magnitude of $10.38\pm0.02$ \citep{cutri03} situates within the operation range of the pyramid wavefront sensor that performs wavefront sensing in $H$-band (${\sim}1.63~\mu$m; \citealp{bond20}) at the Keck II telescope. The single integration time is $0.5$~s, and each exposure frame comprises $60$ integrations. During the observation, we obtained $105$ frames that cover a parallactic angle change of $97\fdg5$ to perform angular differential imaging (ADI). The total on-target integration time is $3150$~s.

The raw exposures of \shortnameobj\ from Keck/NIRC2 require pre- and post-processing to reveal the surroundings that have been overwhelmed by starlight. We followed \citet{xuan18}, which customized the {\tt VIP} package \citep{GomezGonzalez2017}, to perform pre-processing of the data, including flat-fielding, bad-pixel removal, background removal, and image centering. We then performed ADI post-processing using the principal-component-analysis-based Karhunen--Lo\`eve image projection algorithm \citep[KLIP;][]{soummer12, amara12} to remove the central starlight and speckles. In the processing of high-contrast imaging data, ADI and KLIP can introduce reduction artefacts including self-subtraction and over-subtraction, respectively. These artefacts can vary with different reduction parameters, including the number of KLIP components, the rotation angle, etc. To address these artefacts and explore the best limits for our NIRC2 observations, we followed the procedures outlined in \citet{wallack23} and varied these reduction parameters to obtain the deepest contrast limits. Specifically, we generated 5$\sigma$ contrast curves using \texttt{VIP} \citep{GomezGonzalez2017} for different combinations of inner and outer image mask sizes and number of principal components, accounting for the effects of small sample statistics \citep{Mawet2014}. We then compared our achieved contrast across all of the combinations of mask sizes and principal components, and determined the best contrast for an angular separation from the star. We performed the contrast calculation with a 1-pixel step (9.971~mas\footnote{\url{https://github.com/jluastro/nirc2_distortion/wiki}}) to generate an optimal contrast curve for the entire image.

%%%%%%%%%%%%%%%%%%%%%%%%%%%%%%%%%
%%%%%%%%%%%%%%%%%%%%%%%%%%%%%%%%%
\begin{figure*}[!th]
\centering
    \includegraphics[width=\textwidth]{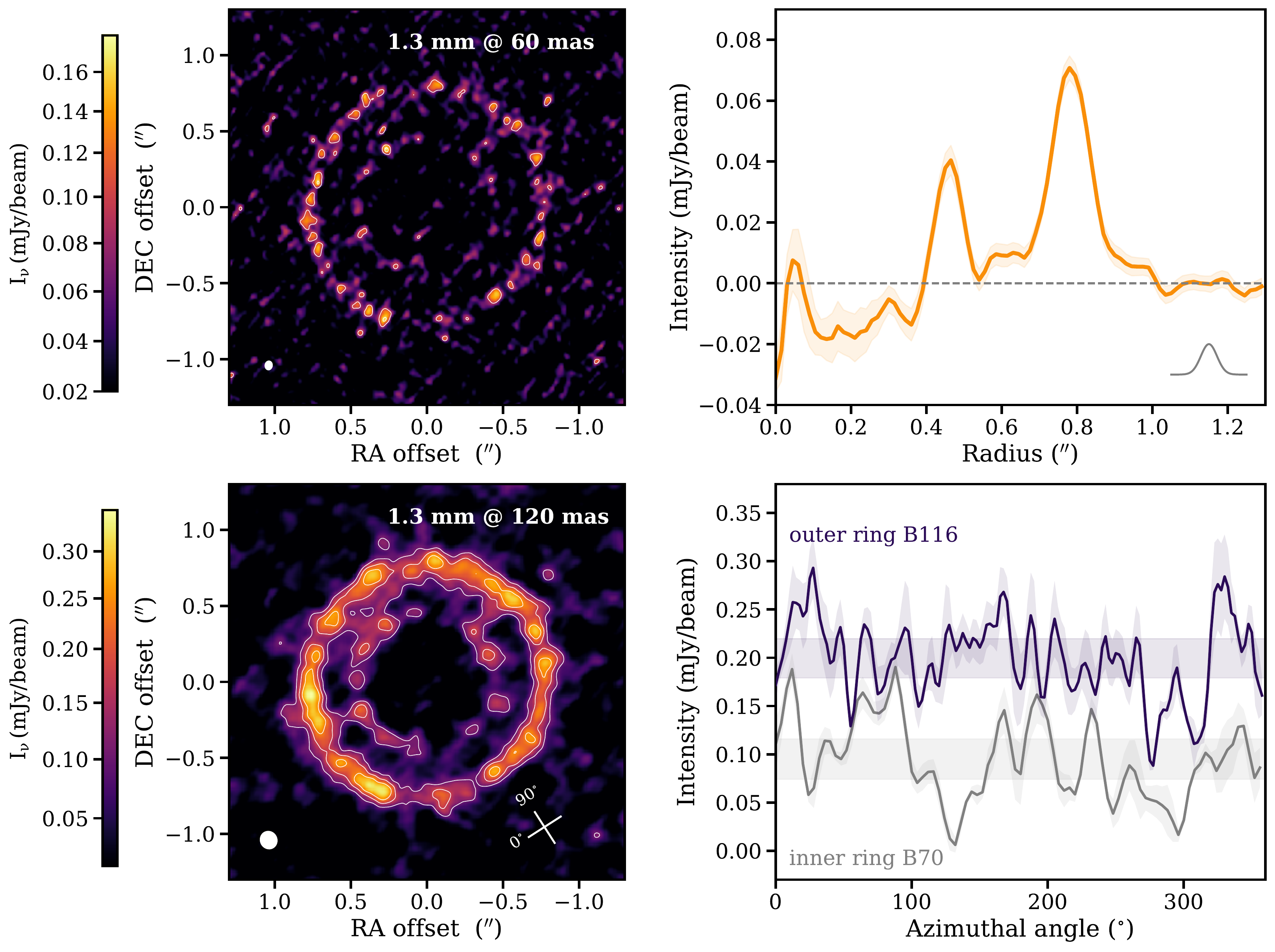}
\caption{{\bf \textit{Top Left:}} Continuum emission image of the \shortnameobj\ disk at 1.3\,mm with a beam size of $0\farcs06\times0\farcs05$. The synthesized beam shape is shown in the bottom-left corner of the panel. White contours are at levels of 3,5$\sigma$.
{\bf \textit{Top Right:}} Azimuthally-averaged radial intensity profile based on image to the left (after deprojection). The Gaussian profile in the bottom-right corner shows the FWHM of the synthesized beam. {\bf \textit{Bottom Left:}} Continuum emission image at 1.3\,mm with a beam size of $0\farcs12$. White contours are at levels of 3,5,7$\sigma$. The azimuthal angle conversion is shown in the bottom right corner. {\bf \textit{Bottom Right:}} Azimuthal intensity profiles at the two dust ring locations using the image to the left. The horizontal shaded regions mark the 1$\sigma$ scatter of dust emission within the ring, centered at the averaged ring intensity.  \label{fig:cont-maps}}
\end{figure*}
%%%%%%%%%%%%%%%%%%%%%%%%%%%%%%%%%
%%%%%%%%%%%%%%%%%%%%%%%%%%%%%%%%%

%%%%%%%%%%%%%%%%%%%%%%%%%%%%%%%%

\begin{deluxetable*}{cccc|cccc}
%[!t]
\tabletypesize{\scriptsize}
\tablecaption{Dust Disk Model Results\label{tab:dust-disk-model}}
\tablewidth{0pt}
\tablehead{
\colhead{Ring Number} & \colhead{Flux} &\colhead{Ring Location} & \colhead{Ring Width ($\sigma$)} & \colhead{incl.} & \colhead{PA} & \colhead{$\Delta_{\rm RA}$} & \colhead{$\Delta_{\rm Dec}$} \\
\colhead{} & \colhead{(mJy)} &\colhead{($''$)} & \colhead{($''$)} & \colhead{($\degr$)} & \colhead{($\degr$)} &\colhead{($''$)} & \colhead{($''$)}  } 
%\colnumbers
\startdata
\#1(B70) & 3.3$\pm$0.3 &  0.472$\pm$0.004 & 0.038$\pm$0.006 & 16.0$\pm$1.1 & 122.6$\pm$4.3 & 0.003$\pm$0.002 & -0.001$\pm$0.002\\
\#2(B116) & 12.7$\pm$0.3 & 0.782$\pm$0.002 & 0.058$\pm$0.003 &  & & &\\
\enddata
\tablecomments{In the fitting, we assumed the same inclination and position angles, and phase center offsets, for both rings. The disk total flux should be the sum of the two rings. A recent SMA observation suggests a total flux of $20\pm2$\,mJy (Liu, Terada et al., private comm.)}
%\tablerefs{}
\end{deluxetable*}

%%%%%%%%%%%%%%%%%%%%%%%%%%%%%%%%%
%%%%%%%%%%%%%%%%%%%%%%%%%%%%%%%%%
\begin{figure*}[!th]
\centering
    \includegraphics[width=\textwidth, trim=0 30 0 50, clip]{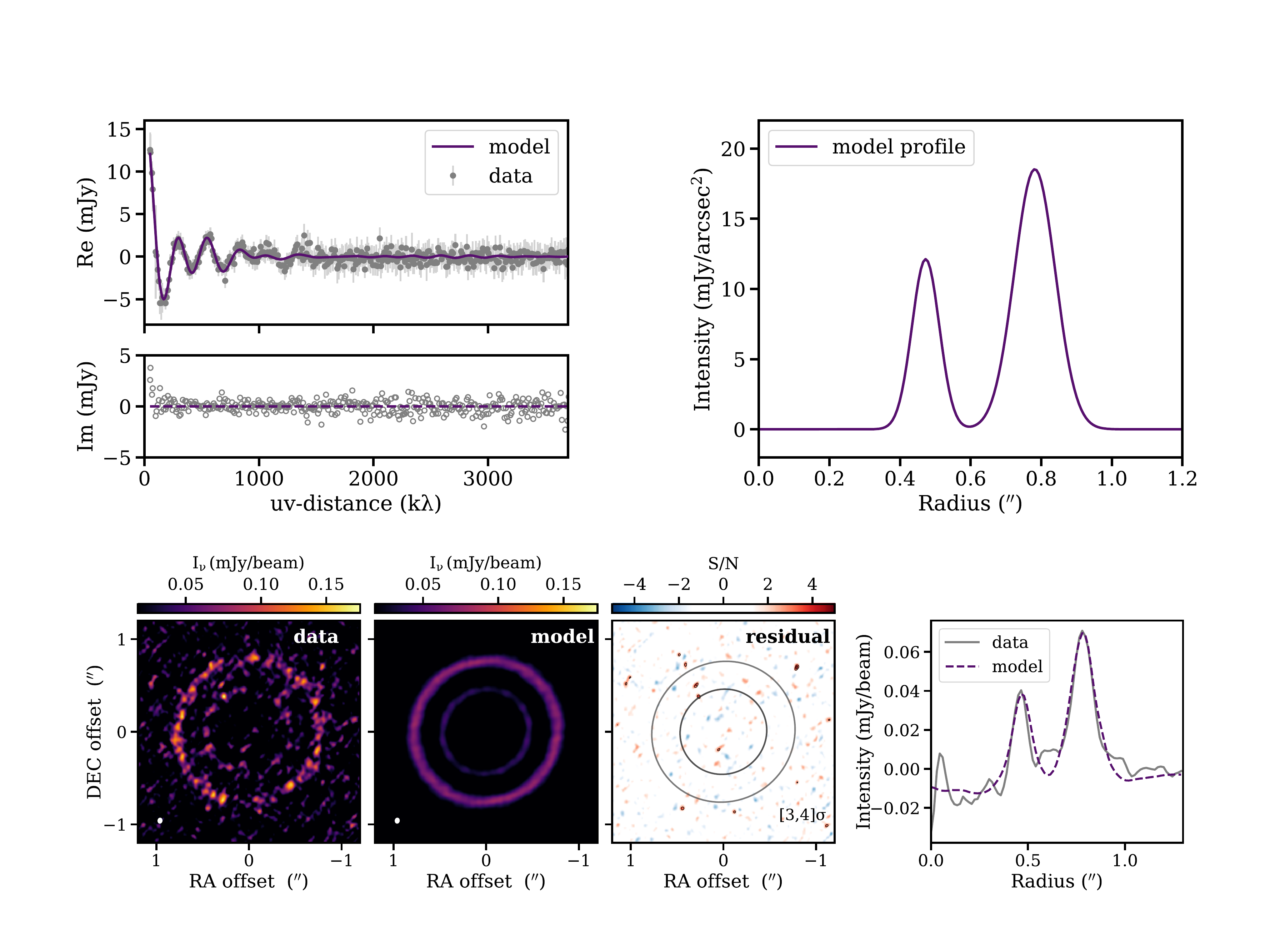} 
\caption{{\bf \textit{Top Left:}} The deprojected and binned visibilities as a function of baseline length for the data and the axisymmetric model; {\bf \textit{Top Right:}} The model radial intensity profile; {\bf \textit{Bottom:}} Comparisons of the data, model, and residual in the image plane, constructed with the same \texttt{tclean} parameters. The two ring locations are marked out in the residual map, with black contours at 3 and 4$\sigma$. The comparison of data and mode radial profiles is shown in the rightmost panel to demonstrate the match in the central depression in the image with the observed $uv$ coverage. \label{fig:cont-uv-model}}
\end{figure*}
%%%%%%%%%%%%%%%%%%%%%%%%%%%%%%%%%
%%%%%%%%%%%%%%%%%%%%%%%%%%%%%%%%%

\section{Results} \label{sec:result}

\subsection{A Disk with Two Dust Rings} \label{sec:dust}
The 1.3\,mm continuum image of the \shortnameobj\ disk at a resolution of $0\farcs06$ is shown in the upper left panel of Figure~\ref{fig:cont-maps}. As the peak emission is only detected at ${\sim}6\sigma$ level, a radial intensity profile is created by averaging the emission along the azimuthal direction to boost the significance of faint emission features (displayed in Figure~\ref{fig:cont-maps} alongside the image). The image deprojection adopts an inclination angle of 16\fdg0 and position angle of 122\fdg7 (derived from visibility modeling below). All of the emission is concentrated in two narrow dust rings, peaking at $0\farcs47$ and $0\farcs78$ (70 and 116\,au, hereafter B70 and B116), with the inner ring narrower and fainter than the outer one. 
Both rings are optically thin, with $\tau \sim$\,0.03--0.04, assuming a dust temperature of 15--20\,K (e.g., \citealt{Andrews2013}), in contrast to many dust rings appear to have optical depths of $\sim$0.6 but are probably optically thick with scattering \citep{Huang2018_ring}. The wide gap separating the two rings is nearly empty, in which the average emission does not exceed 3$\times$ the local noise level, where the noise level is estimated as the 1$\sigma$ scatter along the whole azimuth divided by the square root of beam numbers across the region and shown as the shaded region in Figure~\ref{fig:cont-maps}.

A better visualization of the morphology of dust emission is provided by the image with a coarser resolution of 0$\farcs$12 (lower left panel of Figure~\ref{fig:cont-maps}).  Some dusty filaments (at $\sim$3$\sigma$ level) seem to emerge and connect the inner and outer rings, while the low surface brightness of the ring emission suggests that these features are likely due to imaging artifacts (but see a claimed dusty filament across the gap in HD 135344B, \citealt{Casassus2021}). Likewise, the clumpiness of the dust rings could be attributed to the low sensitivity of the observations and imperfect \textit{uv}-sampling. To better demonstrate any azimuthal emission variations, we created intensity profiles along the azimuthal direction by averaging emission within radial annulus of 0$\farcs$44--0$\farcs$52 and 0$\farcs$72--0$\farcs$82 for B70 and B116, respectively. Around the azimuthal angle of 300$\degr$ (see the azimuthal angle conversion in Figure~\ref{fig:cont-maps}), both rings exhibit a drop of dust emission by a factor of two compared with the averaged ring emission.
It is unclear if the faintness of this azimuth in both rings shares a common origin (e.g., inner disk shadowing, though the inner disk of \shortnameobj\, lacks large grains that can emit predominantly at millimeter wavelength), as such a deficit of dust emission is also seen around 140$\degr$ but only in the B70 ring. Small-scale azimuthal asymmetries in dust rings have been predicted in hydrodynamic simulations from a number of instabilities (e.g., \citealt{HuangP2020, Bi2022, Lehmann2022}). Observational evidence of such instabilities would however require deep integration for high-significance detection of asymmetries in dust rings.

% faint inner ring
% horseshoe ring or planet within the gap staves the inner disk 
%2. uv-plane fitting \\
%- clumpiness and any azimuthal asymmetries \\ 
%- ring width  \\ 
%- deviation from the concentric geometry \\
To quantify the dust emission morphology and disk geometry, we performed model fitting in the visibility domain. We adopted an axisymmetric model with two Gaussian rings, representing the two peaks in the radial profile, which can be expressed as 
\begin{equation} \label{eq1}
I(r) = A_{\rm 1} \exp \left[ -\frac{ (r- R_{\rm 1} )^{2} }{2  \sigma_{\rm 1} ^{2} } \right] + A_{\rm 2} \exp \left[ -\frac{ (r- R_{\rm 2} )^{2} }{2  \sigma_{\rm 2} ^{2} } \right],
\end{equation} 
where $A_{i}$, $R_{i}$, and $\sigma_{i}$ are the amplitude, peak location, and width for individual rings, respectively. Four additional disk geometry parameters (inclination $i$, position angle PA, and offsets from the phase center $\Delta_{\rm RA}$ and $\Delta_{\rm Dec}$) were also included in the fitting.
The model visibilities are then calculated through Hankel transformation \citep{Pearson1999} and sampled at the same observed spatial frequencies. 
The comparison of model and data visibilities uses a Gaussian likelihood $\mathcal{L}\propto \exp(-\chi^2/2)$, where $\chi^{2}= \sum \left| V_{\mathrm{obs}}(u_{k}, v_{k})-V_{\mathrm{mod}}(u_{k}, v_{k}) \right|^{2} w_{k}$, with $w_{k}$ as the observed visibility weights. We assumed uniform priors for the fitted parameters and explored the parameter space with 50 walkers and 5000 steps using a Markov Chain Monte Carlo (MCMC) method (\texttt{emcee}, \citealt{ForemanMackey2013}). 
As the number of steps largely exceeds the autocorrelation time, which is on the order of 100 steps, the MCMC chains are believed to have reached a stationary state.\footnote{We note that other approaches, for example, the rank-normalized split-R diagnostic \citep{Vehtari2021}, have been suggested to check the MCMC convergence.}
The final adopted parameters are summarized in Table~\ref{tab:dust-disk-model}, as the median values of the posterior distributions (computed from the last 1000 steps of the chains), with uncertainties estimated from the 16th and 84th percentiles.

Figure~\ref{fig:cont-uv-model} compares the best-fit model to the data. 
Our adopted model reproduces the real part of the deprojected and azimuthally averaged data visibilities reasonably well. The imaginary part should be clustered around zero for axisymmetric emission; the non-zero fluxes in the short baselines suggest the presence of some non-axisymmetric structures on large scales, which our observations may not capture very well (the maximum recoverable scale based on the fifth percentile of the baseline lengths is only 0$\farcs$6). 
The model and residual images, created with the same \texttt{tclean} parameters as the data image, are also shown in Figure~\ref{fig:cont-uv-model}. No residual emission exceeds 5$\sigma$. Interestingly, there is some 4$\sigma$ residual emission within the dust gap (between B70 and B116), as part of the aforementioned dust filament. As there is other residual emission at similar detection significance, confirming that the emission within the gap is real would require deeper observations. 
Similar to the data image, the model image also exhibits negative fluxes in the inner cavity (see bottom right panel in Figure~\ref{fig:cont-uv-model}), suggesting that this feature is likely related to the sparse sampling in the $uv$-space among the short baseline range. The clumpiness in the rings is probably attributed to both the sparse $uv$ coverage (see the model image) and noise. 

% concentric  

The inner ring B70 has a Gaussian width ($\sigma$) of 38\,mas (FWHM of 90\,mas), $\sim$40\% narrower than B116. The integrated flux for each ring is also listed in Table~\ref{tab:dust-disk-model}; the inner ring is a factor of $\sim$4 fainter (in total flux). Assuming a dust temperature of 20\,K\footnote{Here we adopt the widely used dust temperature of 20\,K, but disks around lower mass stars could be cooler. Dust mass will be higher by $\sim50\%$ if assuming 15\,K.} and the DSHARP opacity of $\kappa_{\rm 1.3\,mm}$=1.9\,cm$^2$\,g\,$^{-1}$ for a maximum grain size of 1\,mm\footnote{This is a reasonable assumption based on the dust fragmentation barrier (see Eq. (8) in \citealt{Pinilla2020}). A large source of uncertainty in dust opacity also relies on the unknown properties of grain composition.} \citep{Birnstiel2018}, we calculate the dust masses of 2.5$\pm$0.2 and 9.6$\pm$0.2\,$M_{\earth}$ for B70 and B116, respectively. 
The continuum disk size estimated from the model intensity profile is 0$\farcs$85$\pm0\farcs003$ ($126.4\pm4.5$\,au) when adopting the effective disk size definition at 90\% fractional luminosity ( \citealt{Tripathi2017}, or 0$\farcs$80$\pm0\farcs002$ at 68\%, slightly beyond the peak of the outer ring).

% total flux in each ring, dust mass
% ring properties 

%%%%%%%%%%%%%%%%%%%%%%%%%%%%%%%%%
\begin{figure}[!t]
\centering
    \includegraphics[width=0.45\textwidth, trim=0 0 0 0, clip]{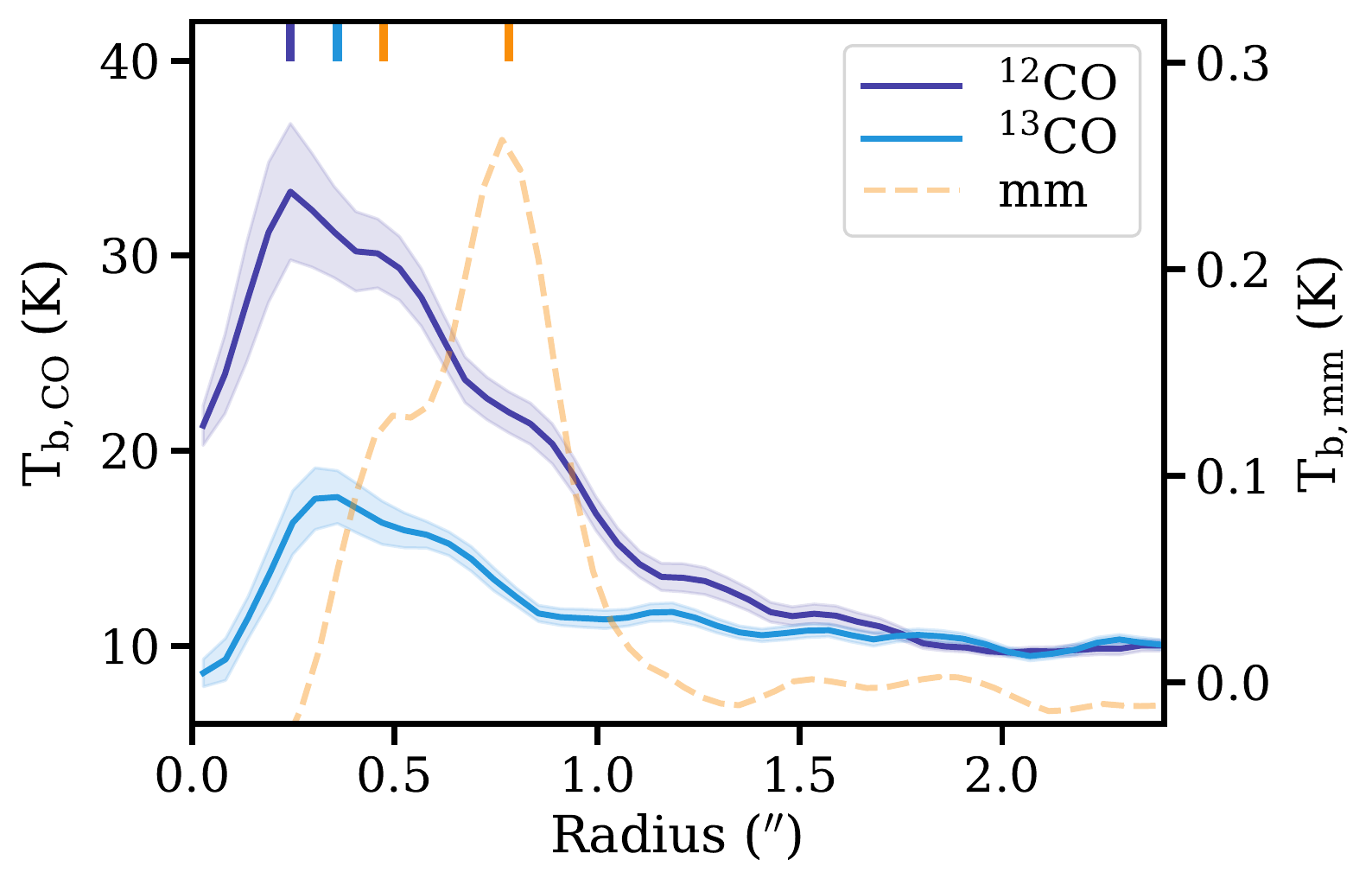}
\caption{Azimuthally averaged radial intensity profiles for CO lines from the deprojected peak intensity maps. The shaded regions mark the 1$\sigma$ scatter at each radial bin divided by the square root of the number of beams across the bin. The continuum emission profile extracted from an image with the same beam size as the line data is shown for comparison. The color ticks at the top of the figure denote the peak emission location for each tracer. 
  \label{fig:cont_co}}
\end{figure}

%%%%%%%%%%%%%%%%%%%%%%%%%%%%%%%%%

\subsection{Dust versus Gas Distributions} \label{sec:cont_co_comp}
%image + radial profile \\
%check if gas cavity is observed in the empty dust cavity 
Three CO isotopologue lines were included in our observations. Emission from $^{12}$CO and $^{13}$CO $J=2-1$ lines is clearly seen in channel maps within a narrow velocity range of 5--8\,km\,s$^{-1}$ (Figure~\ref{fig:co_maps}), mostly tracing the Keplerian disk rotation, though the central channels (around the systemic velocity of $\sim$7\,km\,s$^{-1}$) are cloud contaminated. The integrated line fluxes are 1.01$\pm$0.12 and 0.55$\pm$0.11\,Jy\,km\,s$^{-1}$ for $^{12}$CO and $^{13}$CO lines, respectively, estimated within a circular aperture in radius of 2$\farcs$0 over the velocity range of 5--8\,km\,s$^{-1}$. The uncertainty is measured as the standard deviation of line fluxes from 50 randomly distributed apertures with the same extraction area and velocity range outside the disk emission region. For the C$^{18}$O line, we obtain a line flux upper limit ($3\sigma$) of 0.2\,Jy\,km\,s$^{-1}$.

Figure~\ref{fig:cont_co} compares the radial profiles of CO gas and millimeter dust emission. The line radial profiles were extracted from the peak intensity maps (8th moment) adopting the same disk geometry parameters as the continuum emission for image deprojection\footnote{Radial profiles derived from moment 0 maps show similar variations.}. Within the dust cavity, line emission is clearly detected, but both CO lines exhibit a drop in emission towards the inner region. The identified peak emission locations are 0$\farcs$24 and 0$\farcs$36 for the $^{12}$CO and $^{13}$CO lines, respectively, though the gas cavity sizes may be affected by foreground cloud absorption. 
CO emission is confined to within 2$\farcs$0 from the disk center, resulting in a low CO-to-mm disk size ratio (about 2, see \citealt{Long2022} for this size ratio in a collection of disks). 
As the maximum recoverable scale in this observation is only 0$\farcs$6 (based on the fifth percentile of the baseline lengths), faint line emission on larger scales might be filtered out.

\section{Discussion} \label{sec:diss}

\subsection{Context with Other Disks} 
\begin{figure}[!t]
\centering
    \includegraphics[width=0.45\textwidth, trim=0 10 0 40, clip]{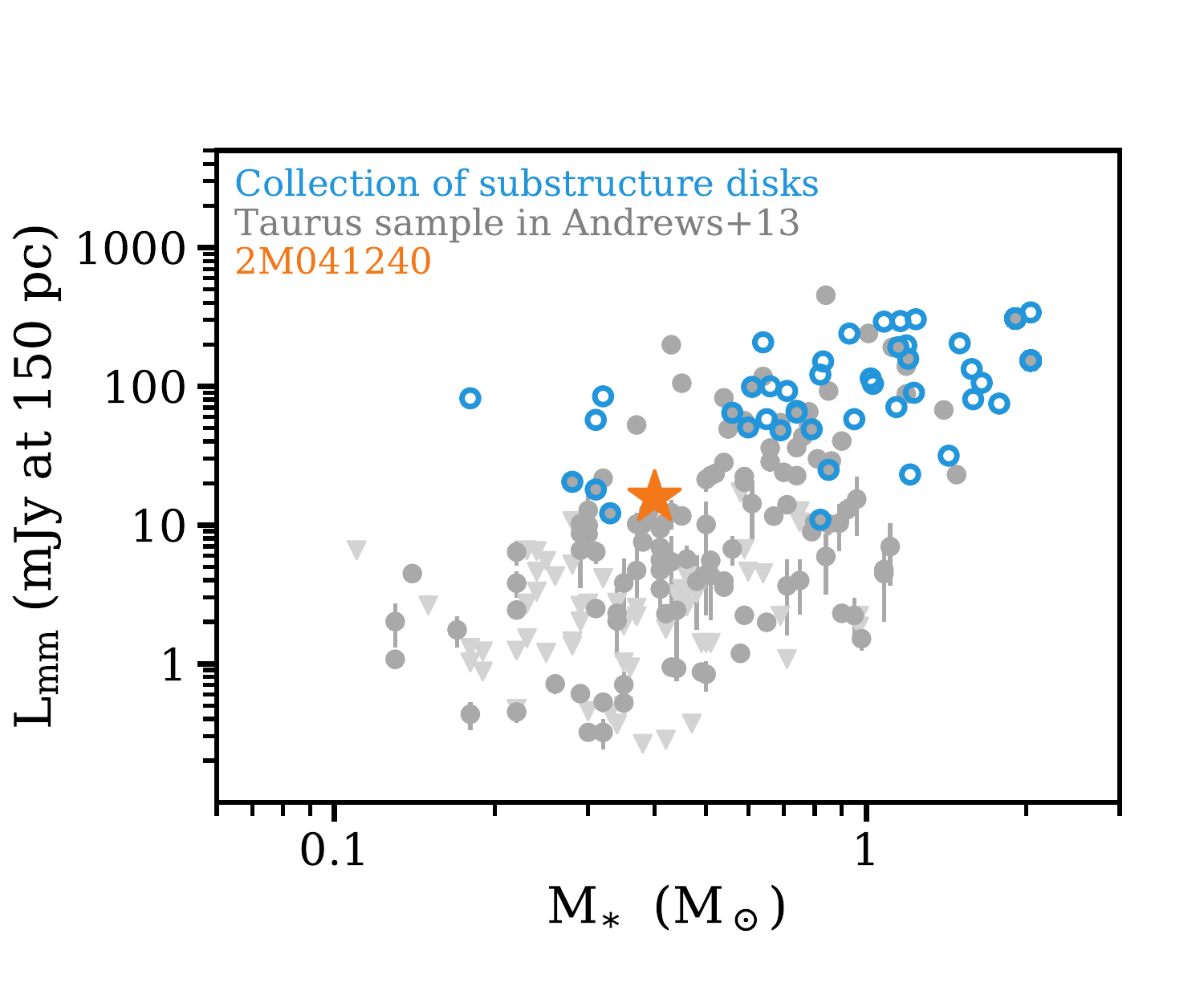} \\
    \includegraphics[width=0.45\textwidth, trim=0 10 0 40, clip]{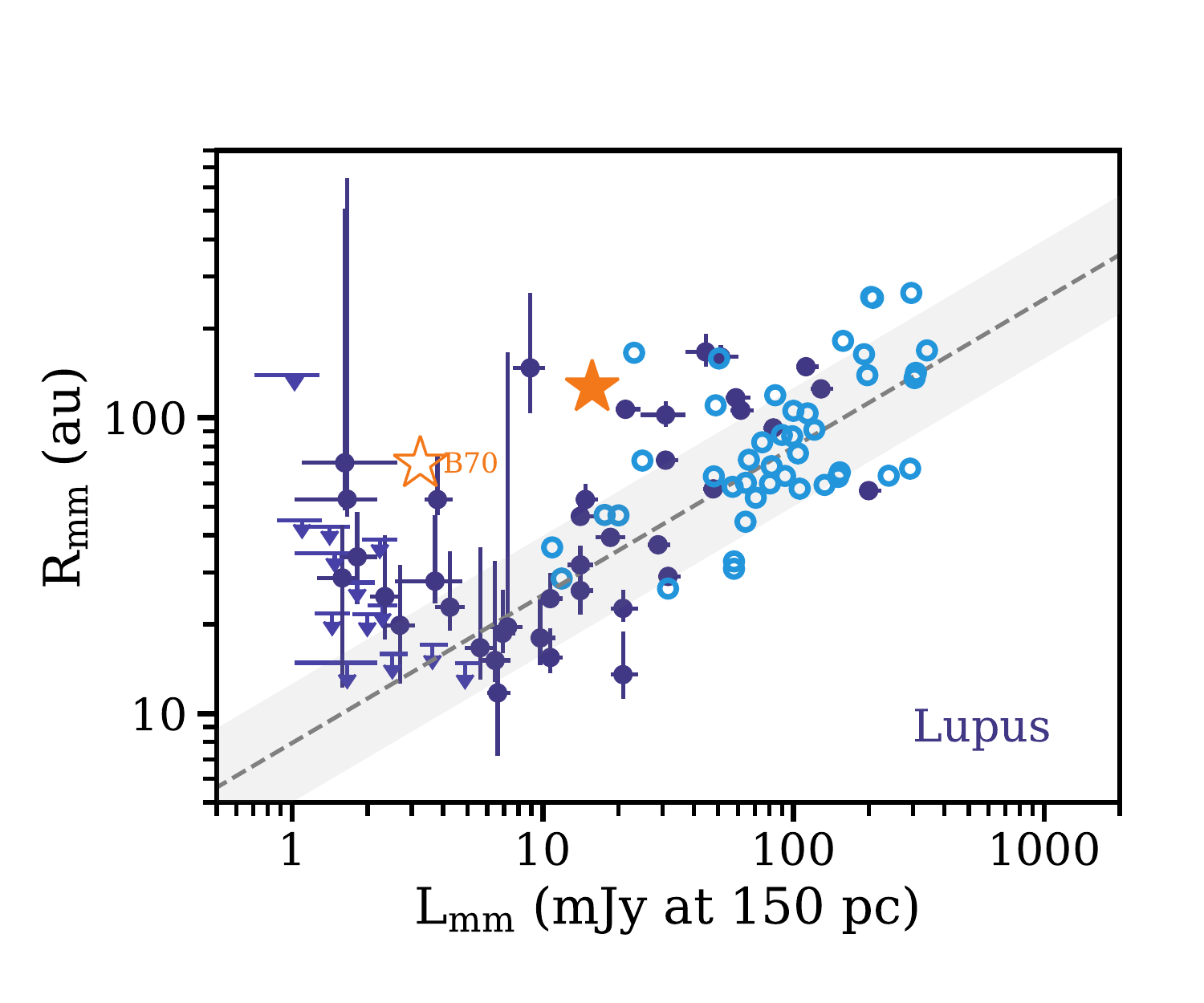} \\
\caption{\textbf{\textit{Top:}} The correlation between stellar mass and disk luminosity at 1.3\,mm for \shortnameobj\, (orange), other Taurus members (grey), and a collection of disks with substructure detection (blue, see references in the main text); \textbf{\textit{Bottom:}} The correlation between disk millimeter luminosity and (90\% fractional) radius. The solid orange star marks the disk of \shortnameobj\,, and the open orange star indicates the inner ring B70. The Lupus sample (in purple), as well as the scaling relation (dashed line) from \citet{Andrews2018_Lmm} are shown for comparison. The lupus sample is adopted for its better coverage at the faint end. The fainter and more extended disk in Lupus than our target is J16090141-3925119, which also surrounds an M4 star. Based on the ALMA image in \citet{Ansdell2016, Ansdell2018}, it remains unclear if it is a disk with a large cavity or a close binary system. See online materials of data used for making the plots.
\label{fig:comp}}
\end{figure}

%put this target into the context of Taurus disk population and other resolved disks with substructures.
Previous disk population studies have revealed a general trend that more massive stars host brighter disks, though associated with significant scatter (e.g., \citealt{Andrews2013, Ansdell2016, Pascucci2016}). The top panel of Figure~\ref{fig:comp} compares \shortnameobj\ with other Taurus members in the $M_{*}-L_{\rm mm}$ plane, where millimeter fluxes for other Taurus disks are taken from \citet{Andrews2013} and updated with new ALMA measurements\footnote{If new measurements are only available at 0.89\,mm, we scaled them to 1.3\,mm using a spectral index of 2.2 \citep{Andrews2020}.} when available \citep{Ward-Duong2018,Akeson2019, Long2019}. We also recalculate the stellar masses using the spot models (with 50\% spot coverage) of \citet{somers20} to be consistent with our adoption of 0.4\,$M_{\odot}$ for \shortnameobj. 
The disk around \shortnameobj\ falls within the top 10\% of millimeter brightness in the group of mid-M-dwarf disks in Taurus (disks around stars of 0.3--0.5\,$M_{\odot}$), a factor of 5 brighter than the median $L_{\rm mm}$ ($\sim$3\,mJy) in this group.   
However, when compared to other disks with known dust substructures in Taurus \citep{Long2018, Kurtovic2021} and other young clusters \citep{Huang2018_ring, Cieza2021}, the disk around \shortnameobj\ lies in the very faint end (Figure~\ref{fig:comp}). 
In particular, we note that the three disks around very low mass stars (M5) with substructures in \citet{Kurtovic2021} have comparable $L_{\rm mm}$ as \shortnameobj. 
As these early high-resolution disk surveys have often selected bright disks, the detection of dust rings in \shortnameobj\ suggests that substructures might also be prevalent in faint disks around low-mass stars, emphasizing the potentially ubiquitous nature of disk substructures\footnote{Substructures have also been found in disks around brown dwarfs (e.g., ISO-Oph 2B, \citealt{Gonzalez-Ruilova2020})}. Systematic high-resolution millimeter imaging surveys of faint disks are needed to determine if substructure is common.

Disks with large inner cavities are typically brighter and have a much shallower $M_{*}-L_{\rm mm}$ relation (equivalent to $M_{*}-M_{\rm d}$ used in \citealt{Pinilla2018}) than that from the full disk populations. With an inner cavity of $\sim70\,$au, the \shortnameobj\ disk deviates significantly (fainter by $\sim 0.5\,$dex) from the relation reported in \citet{Pinilla2018} for inner cavity disks. Dust rings are widely interpreted as pressure bumps that trap particles locally and eliminate the loss of particles from radial drift. In an attempt to understand the observed stellar mass and disk brightness correlations, \citet{Pinilla2020} found that the correlations established from the full disk sample in a number of star-forming regions can be well reproduced by dust evolution and particle trapping models, despite a discrepancy with the sample of bright inner cavity disks. In this model framework, the predicted lower millimeter fluxes are due to the combined effects of high optical depth and efficient grain growth to larger sizes in dust rings (beyond 10\,cm, ``invisible" at millimeter wavelengths, \citealt{Pinilla2020}).

The grain growth effect is more prominent in disks around lower mass stars because of the higher fragmentation barrier that can result in larger grain sizes. Since the rings of \shortnameobj\, have low optical depth (see Section~\ref{sec:dust}), the rapid formation of large grains that have negligible contributions to millimeter fluxes serves as a plausible explanation. 
Given the similar radial scale to cold ExoKuiper belts in debris disks, these dust rings might also be progenitors of old debris rings sustained by the collision of planetesimals formed within these high density regions \citep{Marino2022arXiv, Najita2022}.
On the contrary, those bright disks with inner cavities that depart from the main population, especially those few around low mass hosts ($M_{*}<1\,M_{\odot}$) that drive the $M_{*}-L_{\rm mm}$ relation, might be outliers where grain growth was inhibited within pressure bumps, if somehow grains did not fragment efficiently inside the pressure bumps due to the dust properties (such as composition) or because bouncing stops the effective growth.

Though \shortnameobj\, is an average system in the $M_{*}-L_{\rm mm}$ plane, its dust distribution is far too extended for its given millimeter luminosity, thus sitting high above the well-established scaling relation between $R_{\rm mm}$ and $L_{\rm mm}$ (see the bottom panel of Figure~\ref{fig:comp}, \citealt{Andrews2018_Lmm, Hendler2020}, as a $3\sigma$ outlier considering the derived scatter of 0.2\,dex). 
With a dust disk size of 126\,au, the \shortnameobj\, disk is larger than $\sim90\%$ of all Lupus disks (adopting the $R_{\rm 90}$ results from \citealt{Andrews2018_Lmm}; the Lupus sample is used for comparison due to its higher completeness of size measurement). Among the ring disks with similar $L_{\rm mm}$, the large disk radii of \shortnameobj\, is simply due to the presence of a disk ring at large distance of 110\,au. 
The faint $L_{\rm mm}$ of \shortnameobj\,, compared to other ring disks with similar $R_{\rm mm}$, could be explained as enhanced grain growth in dust rings (as discussed above) and/or the absence of an inner dust disk that might get lost through radial drift with the lack of material supply from the outer region due to pressure trapping. 
It is worth pointing out that the inner ring of B70 alone also stands out as an outlier in the $M_{*}-L_{\rm mm}$ plane, which may suggest that the two rings share a common origin. 
A number of disks in the Lupus star-forming region show similar deviations in the $M_{*}-L_{\rm mm}$ plane at the fainter end (see Figure~\ref{fig:comp}), which are also found to surround M-dwarfs. We suspect that there might be a population of disks around low mass stars that were born large and somehow built a pressure bump early at large disk radii, then evolved along a different path than the disks on the main $M_{*}-L_{\rm mm}$ trend, as shown by \citet{Zormpas2022}.

%morphology: 
%fainter inner ring vs. brighter outer ring 
Large inner cavities were the first type of substructure revealed from initial sub-arcsec resolution disk observations (e.g., \citealt{Andrews2011}). Recently, with the improved spatial resolution of ALMA, the inner cavities are known to be accompanied by dust rings with varying numbers (e.g., \citealt{Loomis2017, Perez2019, HuangJ2020}), but in very rare cases the inner ring is much fainter than the outer one(s). The LkCa~15 disk stands out as such an example, for which \citet{Long2022_lkca15} proposed that the faint inner ring might trace a planet horseshoe orbit with preferential dust accumulation around the two Lagrangian points. Given the lack of such a characteristic feature in the B70 ring of \shortnameobj, this scenario becomes hard to assess, though the total millimeter fluxes in both rings (B70 ring here and horseshoe ring in LkCa~15) are quite comparable. 
Alternatively, \citet{Facchini2020} suggested that the (three-)dust ring configuration in the LkCa~15 disk can be reproduced when considering a vertically isothermal disk with the planet placed in the middle ring. As \shortnameobj\,only contains two rings, it is unclear if these simulations with such an equation of state are applicable to \shortnameobj.

The brightness difference between the two dust rings may also reflect their different dust evolution. If there is only a \textit{weak} pressure bump at B70 (in contrast to a strong one at B116), then millimeter-sized particles may not be effectively trapped and would then quickly migrate inward and get lost onto the central stars. This process might be amplified if the disk has a low gas surface density so that the grains grow larger %can reach higher Stokes number 
and drift faster. This is likely true for \shortnameobj, as its normal dust emission level and large emitting area already imply a low dust surface density. %Meanwhile, the lower fragmentation barrier at closer in radius also suggests grains can grow to even bigger sizes, and no longer contributes to millimeter emission. 

\begin{figure*}[!t]
\centering
    \includegraphics[width=\textwidth]{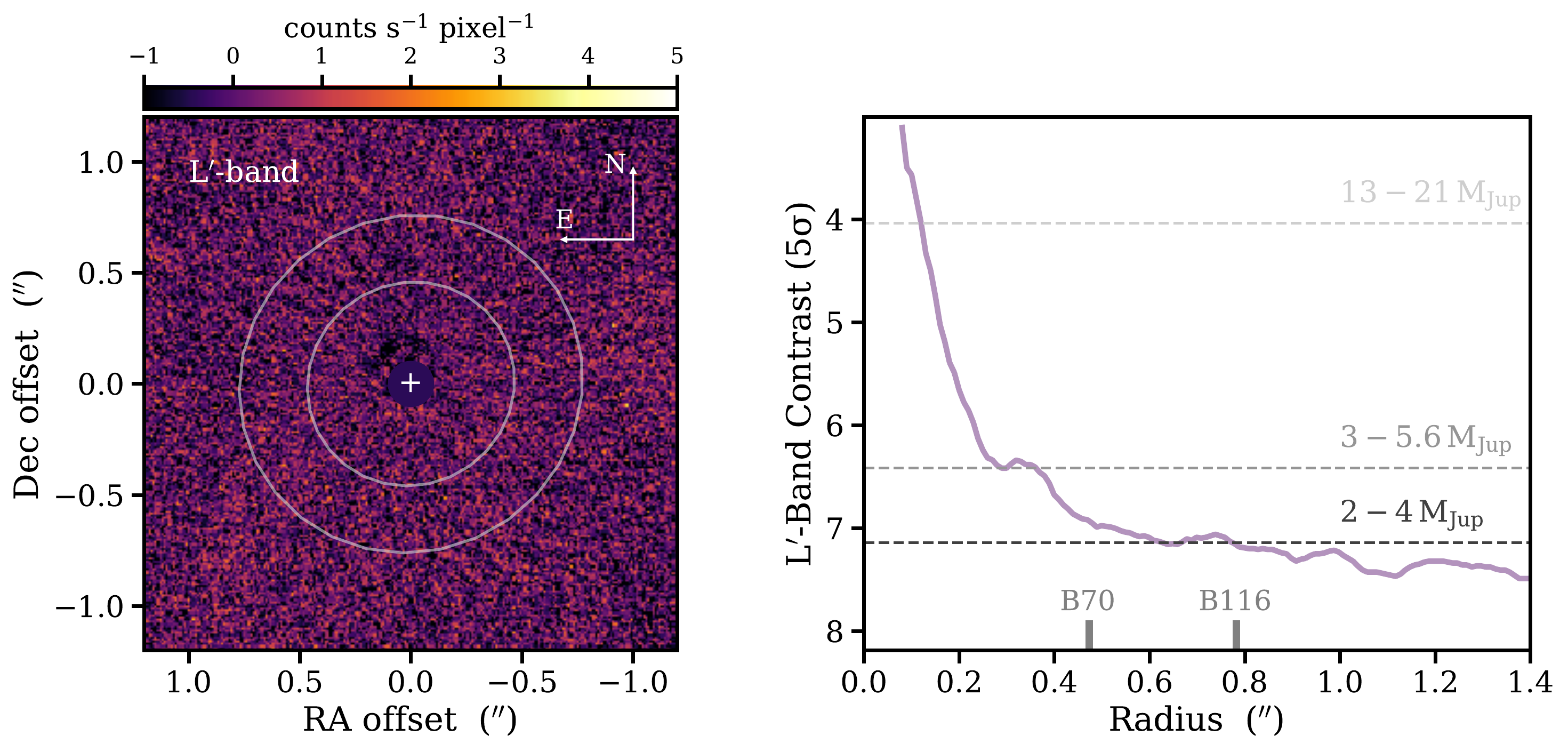}
\caption{\textbf{\textit{Left:}} Keck/NIRC2 image at $L'$-band, reduced with ADI using $10\%$ of the principal components. The two ellipses mark the dust rings identified from ALMA image. We do not identify point sources beyond the $5\sigma$ detection threshold. \textbf{\textit{Right:}} 5$\sigma$ contrast curve. The two dust ring locations are indicated as grey bars at the bottom. The corresponding planet masses at three representative locations (close to the coronagraphic edge, inside the cavity at 0$\farcs$3, and at the dust gap of 0$\farcs$63) are marked out. The mass ranges are given for system ages of 1.6 and 4.5\,Myr. \label{fig:nirc2}}
\end{figure*}

\subsection{Origin of Disk Substructures}
We detect a wide inner cavity and two dust rings peaking at 70 and 116\,au with 1$\sigma$ Gaussian width of 5.6 and 8.5\,au, respectively, in the \shortnameobj\, disk. When considering a passively irradiated flaring disk, the disk mid-plane temperature at a given radius $r$ can be approximated as: 
\begin{equation}
    T_{\mathrm{d}}(r) = \left(\frac{\frac{1}{2}\varphi L_{*}}{4\pi r^2\sigma_{\mathrm{SB}}}\right)^{1/4}
\end{equation}
\citep[e.g.,][]{Chiang1997, Dullemond2001}, where $\varphi$ is the disk flaring angle, $L_{*}$ is the stellar luminosity, and $\sigma_{\mathrm{SB}}$ is the Stefan-Boltzmann constant. Given the low luminosity of the central star, the disk radius corresponding to a mid-plane temperature of 15\,K is well within 50\,au. Therefore, these substructures are unlikely related to the ice lines of major volatile species \citep{Zhang2015, Long2018}. It is worth noticing that ice lines can be very dynamic and difficult to model (e.g., \citealt{Owen2020}), and obtaining a precise mid-plane temperature is challenging. 

The comparison of ring width to local pressure scale height ($h_p$) can help evaluate if grains are trapped in pressure bumps. This is because a long-lived gas pressure bump is supposed to be wider than $h_p$, while the trapped dust can be concentrated in narrower regions \citep{Dullemond2018}.
The disk pressure scale height can be estimated as $h_p$ = $\sqrt{\frac{k_{\rm B}T_{\rm d}r^3}{\mu m_{p}GM_*}}$, where $k_{\rm B}$ is the Boltzmann constant, $m_{p}$ is the proton mass, $G$ is the gravitational constant, and $\mu=$\,2.3 is the mean molecular weight in atomic units.
Using a temperature profile with $\varphi$=0.02 (a conservative choice), the corresponding pressure scale heights at the two ring locations are 5.7 and 10.7\,au, respectively.
While the width (Gaussian $\sigma$) of the inner ring B70 is comparable to $h_p$, the outer ring B116 is narrower than $h_p$ by $\sim25\%$, providing strong evidence of dust trapping operating there.

Local pressure maxima that trap dust grains can occur in many different ways. For example, around the outer edge of a dead zone (induced from magneto-hydrodynamical effects), a pressure trap is created due to the sharp change of disk viscosity (e.g., \citealt{Flock2015, Pinilla2016}). However, such a location is expected to be within 20\,au in disks around M-dwarfs \citep{Delage2022}, thus the dead zone hypothesis alone is not applicable to the \shortnameobj\, disk rings.

The recent study by \citet{Garate2021} proposed that the combined effects of dead zone and X-ray-driven photoevaporation may explain the set of transition disks with wide cavities ($>$20\,au) and considerable mass accretion rates. 
Though photoevaporation models can open a gap with tens of au radius, they often fail to reproduce the observed accretion rates \citep[e.g.,][]{ErcolanoPascucci2017, Picogna2019}.  \shortnameobj\, is actively accreting at a rate of $\sim 1.4\times10^{-9}~M_\odot$ yr$^{-1}$, typical for its stellar mass \citep{alcala17}, which rules out photoevaporation alone as a viable mechanism. One key component of the \citet{Garate2021} model relies on the slower viscous evolution in the dead zone region, which sustains a higher accretion rate on a prolonged timescale. However, the probability of creating a 70\,au cavity with the needed accretion rate is still low based on their model grids, and the outcome highly depends on the assumed $\alpha-$viscosity in the dead zone region. Their models also expect a bright inner disk at millimeter wavelength, which is not seen in our ALMA images.

One intriguing possibility is that one or more planets embedded in the disk may be responsible for the observed dust gaps and rings. The propagation and dissipation of spiral density waves driven by planet--disk interactions redistribute disk material, which also allows gas and small grains to flow through the gaps (e.g., \citealt{Dodson-Robinson2011, Zhu2012}). Figure~\ref{fig:nirc2} presents the results of our search for such dynamical perturbers in the $L'$-band of Keck/NIRC2. For each angular separation from the center, we measure the noise and calculate the signal-to-noise ratio for all pixels; we do not identify any sources above $5\sigma$.
In order to determine the mass upper limits corresponding to our optimal contrast curve, we use the AMES-Cond models \citep{Baraffe2003}, assuming the $L'$ magnitude from the logarithmic interpolation of the WISE $W1$ and $W2$ magnitudes \citep{Marocco2021}, a system age of 4.5\,Myr, and a distance of 148.7 pc. 
Planet mass limits for three representative locations are marked out in the right panel of Figure~\ref{fig:nirc2}. In the middle of the two dust rings (at 0$\farcs$63, 93\,au), our observations reach a contrast level of $\sim$7.1\,mag, corresponding to a planet mass upper limit of $\sim3.9\,M_{\rm Jup}$ (or $\sim2.1\,M_{\rm Jup}$ when adopting an age of 1.6\,Myr). Inside the first dust ring at 0$\farcs$3, we can obtain a mass upper limit of $\sim5.6\,M_{\rm Jup}$ (or $\sim3.1\,M_{\rm Jup}$ at 1.6\,Myr).

Our Keck observations rule out stellar companions on circular orbits within $\sim$10--70\,au, while the stellar multiplicity within 10\,au is unclear. For circumbinary disks, tidal forces tend to carve out an inner disk cavity, whose size scales with the binary semimajor axis, usually by a factor of 2--3 (e.g., \citealt{Artymowicz1994}). Therefore, a stellar companion within 10\,au ($\sim$ the coronagraph size) may not be able to explain our observed wide cavity in \shortnameobj, unless the binary is on a very eccentric orbit. Famous examples include HD142527 and IRAS 04158+2805, where large cavities of $\sim$90 and 185\,au in radii are believed to be carved by highly eccentric binaries at projected separations of $\sim$13 and 25\,au, respectively \citep{Price2018, Ragusa2021}. Both of those cases show high-contrast azimuthal asymmetries in the dust ring, which is not obviously seen in the \shortnameobj\, disk. Interestingly, recent models of the circumbinary disk CS Cha suggest that the inclusion of a planet inside the cavity orbiting around the close binary could damp the cavity eccentricity and largely reduce the dust ring asymmetries \citep{Kurtovic2022}. Thus, we could not safely rule out \shortnameobj\, as a binary system, but a binary alone may not explain the observed disk morphology.

% describe first the inner cavity 
Deficits of CO emission are found within the inner dust cavity, where the $^{13}$CO emission peaks at a radial distance of 0$\farcs$36, closer in than the millimeter dust. 
Assuming that the dust rings were produced by embedded planets, based on the criterion\footnote{A gas gap is opened if the disk satisfies the condition of $\frac{3}{4}\frac{H}{R_H}+\frac{50}{q Re}\lesssim 1$, where $H$, $R_H$, $q$, and $Re$ are disk scale height, planet Hill radius, planet-to-star mass ratio, and Reynolds number as ${r_p}^2\Omega_{p}/\nu$, respectively. } proposed by \citet{Crida2006}, we calculate the required minimum planet mass to open a gap in the gas surface density that creates a local pressure bump and then efficiently traps dust grains. 
If a planet placed at 0$\farcs$3 is responsible for the wide inner cavity, the estimated planet-to-stellar mass ratios are 5.5$\times10^{-4}$ and 1.1$\times10^{-3}$ for $\alpha$-viscosity of $10^{-4}$ and $10^{-3}$, respectively, corresponding to 0.24\,$M_{\rm Jup}$ and 0.48\,$M_{\rm Jup}$ for a 0.4\,$M_{\odot}$ star. 
Planet--disk interaction models often predict spatial segregation of emission peaks between millimeter dust and gas (or small dust, e.g., \citealt{deJuanOvelar2013, Facchini2018}), as seen in our case. 
The spatial difference of emission peaks is expected to increase with planet mass (albeit with dependencies on other disk parameters, including viscosity). Following Eq.~4 in \citet{Facchini2018}, we obtain a planet-to-star mass ratio of 1.2$\times10^{-3}$ based on the ratio of emission peaks ($R_{\rm mm, B70}/R_{\rm 13CO} \sim$ 1.3), in good agreement with the estimates above. We note that the gas cavity size could be overestimated due to cloud contamination, so that the responsible planet might have lower mass. In addition, the models in \citet{Facchini2018} focused on Sun-like stars, and it remains unclear how this peak offset may change in disks around lower mass stars where the conditions for gas--dust coupling may differ.

At the gap location between the two rings, the \citet{Crida2006} criterion expects a planet mass of 0.39--0.78\,$M_{\rm Jup}$. As no clear CO gas gap is identified at the dust gap location, these values may be treated as certain upper limits. Following the method of \citet{Lodato2019} that relates the dust gap width to the planet Hill radius, we obtain a planet mass of 0.21\,$M_{\rm Jup}$ when adopting a scaling factor of 4.5 (larger factors lead to lower masses). Those estimates are well below the upper limits derived from Keck observations. If we take the dust disk mass estimates in Section~\ref{sec:dust} and assume a gas-to-dust mass ratio of 100, the current disk mass is $\sim$3.8\,$M_{\rm Jup}$. Taking into account the disk mass accreted onto the star ($\sim$7.8\,$M_{\rm Jup}$, assuming a constant mass accretion rate within the system age), the formation of those planets would make use of about 10\% of the total initial disk mass. 
Thus from the disk mass budget perspective, this rough estimate suggests that planet-disk interaction is a plausible explanation for the observed gaps and rings in the \shortnameobj\, disk.

The fact that one single planet can create multiple rings observable at millimeter wavelength in a low viscosity disk (e.g., \citealt{Dong2017, Bae2017, Perez2019}) has made it challenging to predict the number of planets in a system; a simple allocation of one planet to one dust gap can be problematic. On the other hand, a wide cavity may host multiple planets (e.g., two giant planets detected within the cavity of PDS~70, \citealt{Keppler2018, Haffert2019}). 
Planet migration at an appropriate speed can also produce double-ring emission features \citep{Meru2019}, which can be tested with spectral index analysis by combining multi-wavelength observations \citep{Nazari2019}. More recently, \citet{Kanagawa2021} further explored how a migrating planet affects dust ring morphology in relatively low viscous disks ($\alpha \sim 10^{-4}$) and found that the initial outer dust ring would gradually broaden and fade away with time if the planet migration timescale is shorter than the gap formation timescale (dust ring will not follow the planet's movement). Following the criterion proposed in \citet{Kanagawa2021}, the B116 ring in the disk of \shortnameobj\, might indicate the initial location of the migrating planet, while its present location is very uncertain, as it highly depends on many other disk properties, in particular disk viscosity.

Though current ground-based AO facilities are only sensitive to planets with a few Jupiter masses, JWST opens a new discovery space. 
To evaluate the prospects for detecting a putative planet in the disk of \shortnameobj\, with JWST/NIRCam using a coronagraph, we simulate the observations using python package \texttt{pynrc} \citep{leis2022}, 
%\footnote{\url{https://github.com/JarronL/pynrc}}, 
consistent with the observational setup and results from \citet{carter22} and \citet{girard22}.
Considering the expected contrast between the planet and the star (plus the inner disk), F444W is the optimal band to search for a gas giant planet.
In a 4,000 sec observation, the $5\sigma$ contrast at $0\farcs6$ from the star is 11 mag ($\sim 5\times10^{-5}$). According to the \citet{linder19} models that are suitable for lower mass planets, this contrast corresponds to $\sim 0.6$ $M_{\rm Jup}$ for a 5 Myr old planet (or $\sim 0.3$ $M_{\rm Jup}$  for 2 Myr), within the range where the inferred planet in the gap might be detectable.  
The contrast at $0\farcs3$ would be $8$ mag, corresponding to the brightness of a 3\,$M_{\rm Jup}$ planet at 5\,Myr,  so the planet within the cavity is likely not detectable.
Most ringed disks are 2-3 mag brighter in the F444W filter, which limits the sensitivity to planets less massive than Jupiter. Compared to many other disks with dust gaps and rings at tens of au radii, the dimmer stellar emission of \shortnameobj\ makes it a promising target for JWST searches for Saturn-mass planets.

\section{Summary} \label{sec:sum}
This paper presented new ALMA Band 6 observations of the disk around a low mass star \obj. We applied parametric Gaussian models to characterize the dust emission morphology in the visibility plane and performed Keck AO observations to explore the cause of the observed disk substructures. Our main findings are summarized as follows: 

\begin{enumerate}

\item With a spatial resolution of $\sim$50\,mas (8\,au), we detected a large dust cavity surrounded by a dust ring at 70\,au, followed by another brighter and wider ring at 116\,au. Both rings are quite narrow, with gaussian widths of 5.6 and 8.5\,au, respectively. The outer ring is narrower than the local pressure scale height, suggesting the presence of pressure bump that traps dust particles.
Though the two dust rings are only detected at low significance (peak brightness of 6--8$\sigma$), the large surface area of emission makes it a normal bright disk for its spectral type. However, its dust disk size (126\,au for $R_{90\%}$) is much larger than disks with similar millimeter luminosity. The \shortnameobj\, disk likely formed large and built the pressure bump at early times that sustained mm-sized grains at large radii.

\item We searched for possible young planets in the disk using Keck/NIRC2 high-contrast imaging observations at $L'$-band (${\sim}3.78~\mu$m) and did not identify any 5$\sigma$ signals. These observations constrain the potential planets below 2--6 $M_{\rm Jup}$ outside $\sim$50\,au, and rule out any substellar mass companions outside $\sim$10\,au. Despite the non-detections, planet--disk interactions remain the most likely explanation for the observed gaps and rings, as we explored the possibilities of a few other mechanisms (including ice lines, dead zone, and photoevaporation wind). Based on the gap properties, we suggest the presence of a $\sim$Saturn-mass planet at $\sim$90\,au, consistent with the constraints set by direct imaging observations.

\item Among the three targeted CO isotopologue lines, both $^{12}$CO and $^{13}$CO line emission is detected, though strongly affected by foreground cloud absorption. Both lines show deficits of emission in the inner disk, peaking inside the first dust ring. Compared with existing models, the cavity size difference between the CO gas and mm dust also points to a $\sim$Saturn-mass planet in the inner disk. 

\end{enumerate}

There is an increasing number of disks around low mass stars that show dust substructures. Substructures are likely ubiquitous in all types of disks. M-dwarf disks with prominent gaps and rings provide favorable conditions for young planet search. On one hand, the increased scale height in disks around lower mass stars requires higher planet-to-star mass ratios to open the gap. On the other hand, the fainter stellar emission allows us to probe lower mass planets for a given contrast. Future JWST observations can potentially detect the predicted $\sim$Saturn-mass planet in the outer disk of \shortnameobj.

\begin{comment}
\begin{figure}[!t]
\centering
    \includegraphics[width=\columnwidth]{Crida_criterion_01Msun_alpha1E4.pdf}
\caption{Crida criterion to identify the minimum mass planet to open a gap in a disk around 2M041240
  \label{fig:crida}}
\end{figure}
%\subsection{The Possible Planet in the Inner Cavity?}
\end{comment}

%a valuable target for understanding the formation and evolution of disk 

\paragraph{Acknowledgments}
%ALMA team's acknowledgement
F.L. thanks Rixin Li and Shangjia Zhang for insightful discussions.
Support for F.L. was provided by NASA through the NASA Hubble Fellowship grant \#HST-HF2-51512.001-A awarded by the Space Telescope Science Institute, which is operated by the Association of Universities for Research in Astronomy, Incorporated, under NASA contract NAS5-26555. F.L. and S.A. acknowledge funding support from the National Aeronautics and Space Administration under grant No.17-XRP17$\_$2-0012 issued through the Exoplanets Research Program.
D.H. is supported by Center for Informatics and Computation in Astronomy (CICA) grant and grant number 110J0353I9 from the Ministry of Education of Taiwan. DH also acknowledges support from the National Science and Technology Council of Taiwan through grant number 111B3005191.
G.J.H and Y.S. are supported by grant 12173003 from the National Natural Science Foundation of China.
G.D.M. acknowledges support from FONDECYT project 11221206, from ANID --- Millennium Science Initiative --- ICN12\_009, and the ANID BASAL project FB210003.
D.J. is supported by NRC Canada and by an NSERC Discovery Grant. 
Y.L. acknowledges the financial support by the Natural Science Foundation of China (Grant No. 11973090).
L.C. was supported by ANID through the Milennium Science Initiative Program (NCN2021\_080) and the FONDECYT Grant 1211656.
E.R. acknowledges financial support from the European Research Council (ERC) under the European Union's Horizon 2020 research and innovation programme (grant agreement No. 864965, PODCAST) 
C.F.M is funded by the European Union (ERC, WANDA, 101039452). Views and opinions expressed are however those of the author(s) only and do not necessarily reflect those of the European Union or the European Research Council Executive Agency. Neither the European Union nor the granting authority can be held responsible for them.
%Bin's acknowledgement from EU
This project has received funding from the European Research Council (ERC) under the European Union's Horizon 2020 research and innovation programme (PROTOPLANETS, grant agreement No. 101002188). 
This research was funded in part by the Gordon and Betty Moore Foundation through grant GBMF8550 to M.~Liu. 
K.W. is supported by NASA through the NASA Hubble Fellowship grant HST- HF2-51472.001-A awarded by the Space Telescope Science Institute, which is operated by the Association of Universities for Research in Astronomy, Incorporated, under NASA contract NAS5-26555.
J.L. acknowledges funding from JWST/NIRCam contract to the University of Arizona, NAS5-02105
%Caltech team's acknowledgement
This research is partially supported by NASA ROSES XRP, award 80NSSC19K0294. Some of the data presented herein were obtained at the W.~M.~Keck Observatory, which is operated as a scientific partnership among the California Institute of Technology, the University of California and the National Aeronautics and Space Administration. The Observatory was made possible by the generous financial support of the W.~M.~Keck Foundation. The authors wish to recognize and acknowledge the very significant cultural role and reverence that the summit of Maunakea has always had within the indigenous Hawaiian community.  We are most fortunate to have the opportunity to conduct observations from this mountain. Part of the computations presented here were conducted in the Resnick High Performance Computing Center, a facility supported by Resnick Sustainability Institute at the California Institute of Technology. 

This paper makes use of the following ALMA data: ADS/JAO.ALMA\#2019.1.00566.S. ALMA is a partnership of ESO (representing its member states), NSF (USA) and NINS (Japan), together with NRC (Canada), MOST and ASIAA (Taiwan), and KASI (Republic of Korea), in cooperation with the Republic of Chile. The Joint ALMA Observatory is operated by ESO, AUI/NRAO and NAOJ. The National Radio Astronomy Observatory is a facility of the National Science Foundation operated under cooperative agreement by Associated Universities, Inc.

\facilities{ALMA, Keck II (NIRC2)}
\software{{\tt emcee} \citep{ForemanMackey2013}, CASA \citep{CASATeam2022PASP}, {\tt pynrc} \citep{leis2022},  {\tt VIP} \citep{GomezGonzalez2017}.}

\appendix

\section{CO channel maps} \label{sec:co_maps}
\begin{figure*}[!h]
\centering
    \includegraphics[width=0.8\textwidth]{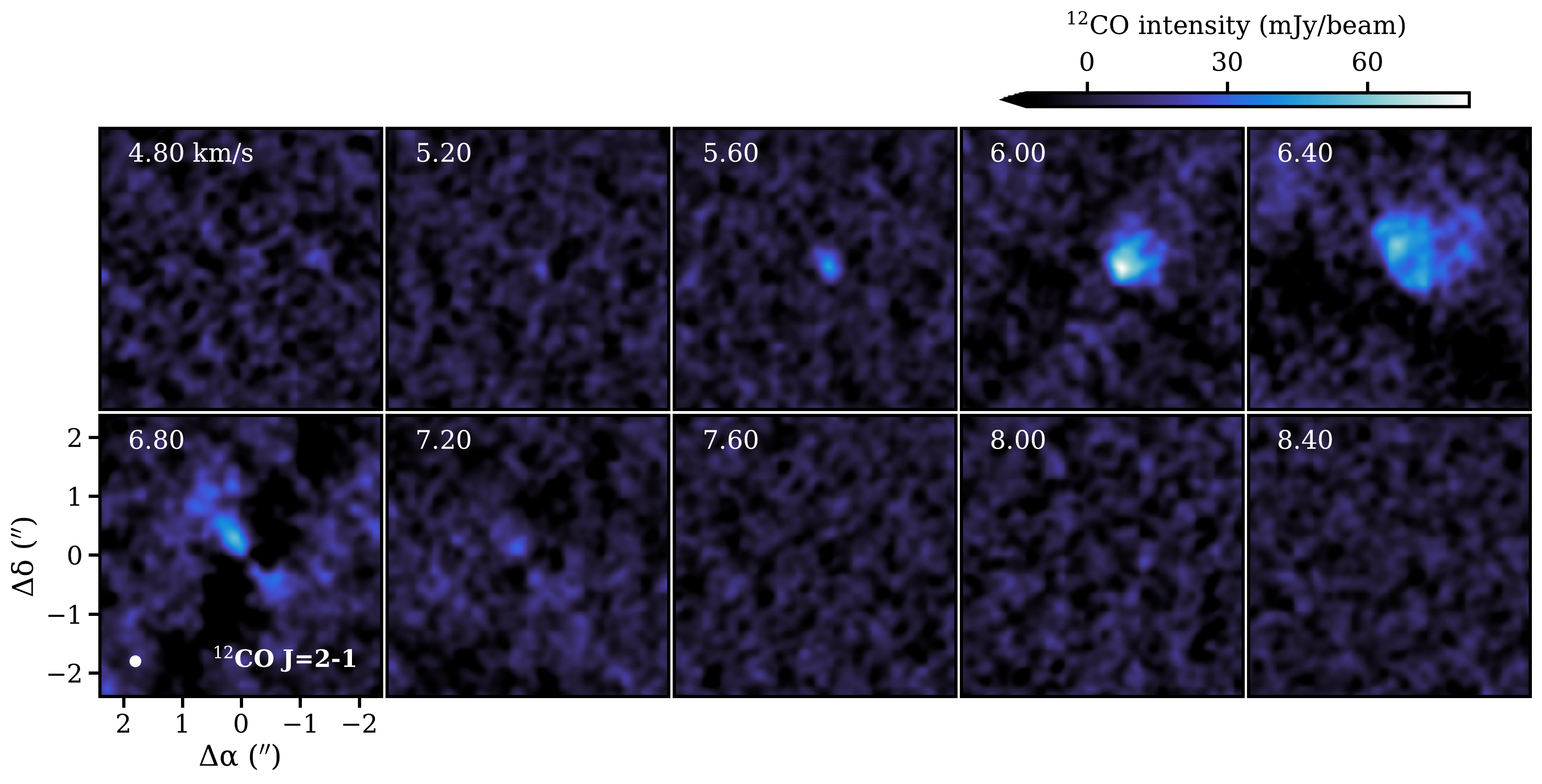} \\
    \includegraphics[width=0.8\textwidth]{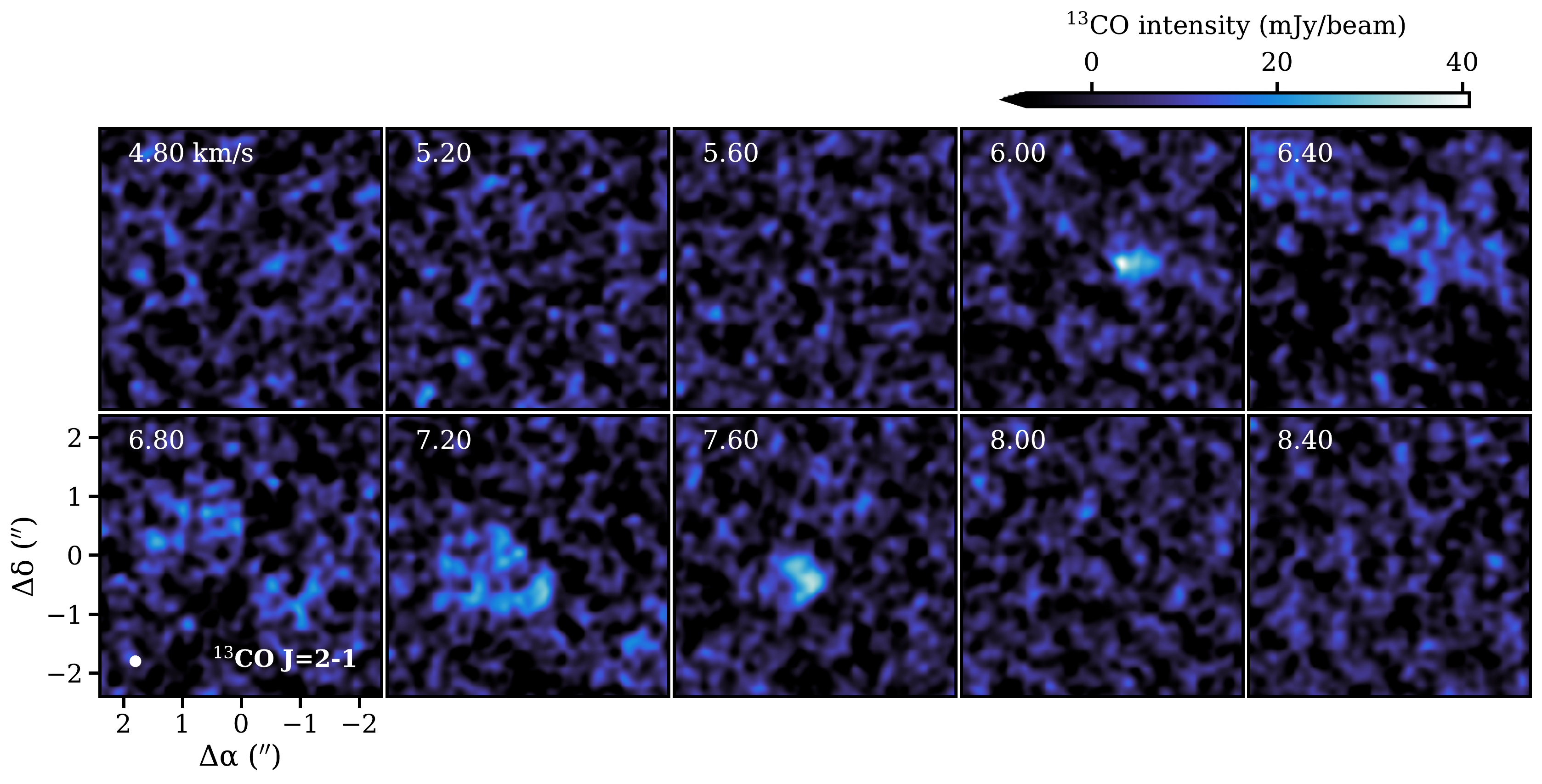} \\
    \includegraphics[width=0.8\textwidth]{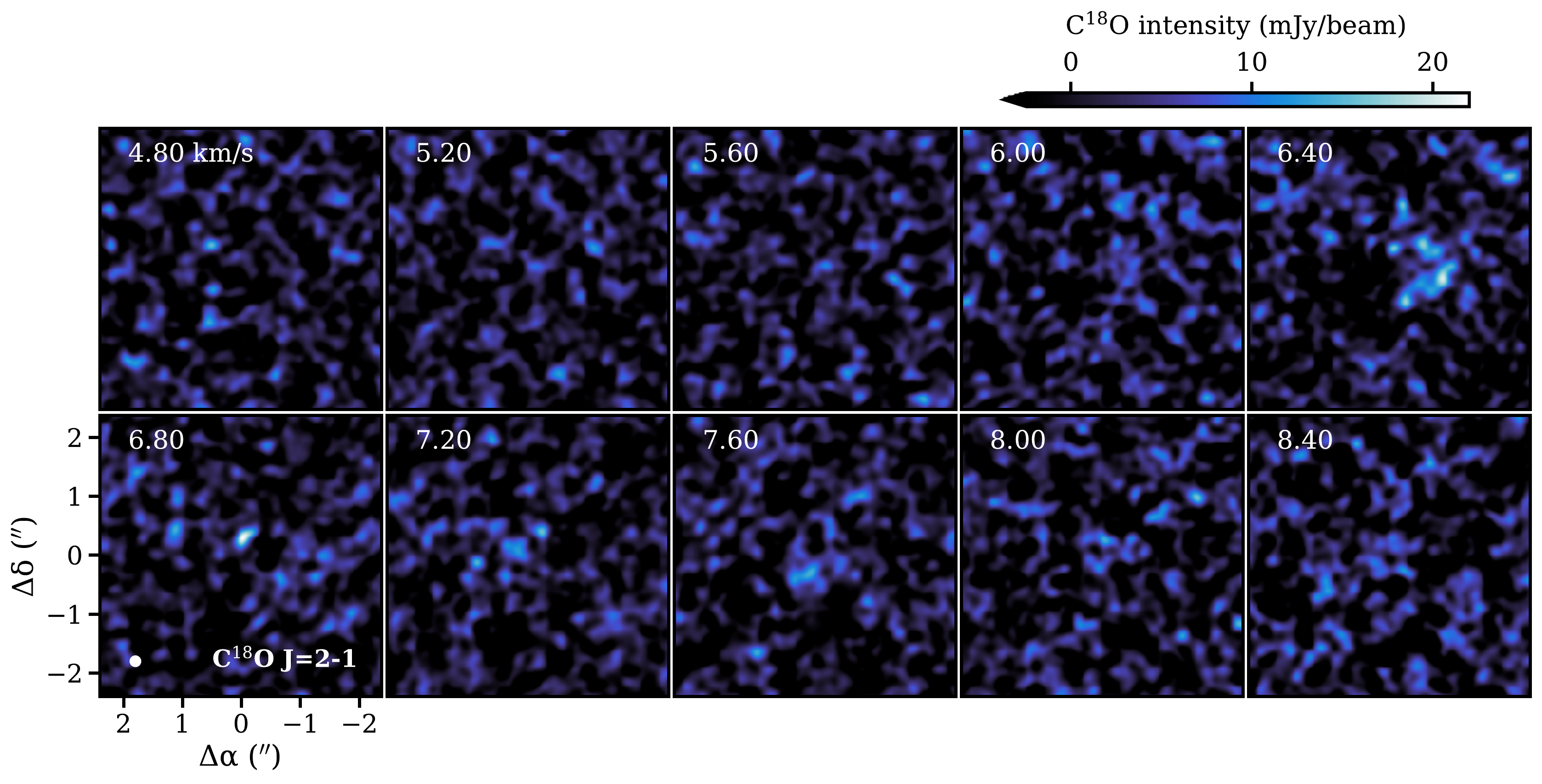} \\
    \caption{The channel maps for line emission of $^{12}$CO (upper) and $^{13}$CO (lower) $J=2-1$ in the \shortnameobj\ disk. The line central channels (around 7\,km\,s$^{-1}$) are affected by foreground cloud contamination. The beam size is shown in the left corner of the first lower panel.  \label{fig:co_maps}}
\end{figure*}

The channel maps for $^{12}$CO and $^{13}$CO $J=2-1$ lines are shown in Figure~\ref{fig:co_maps}. 
A Gaussian uvtaper ($0\farcs2\times0\farcs18$, -60$\degr$) was applied in \texttt{tclean} to achieve a synthesized beam of $0\farcs22\times0\farcs21$ for a better visualization of the faint line emission. Line emission is detected within a velocity range of 5--8\,km\,s$^{-1}$, where the central channels are contaminated by the foreground cloud. For $^{12}$CO, the blue-shifted velocity channels are also heavily contaminated. 

\begin{figure}[!h]
\centering
    \includegraphics[width=0.95\columnwidth]{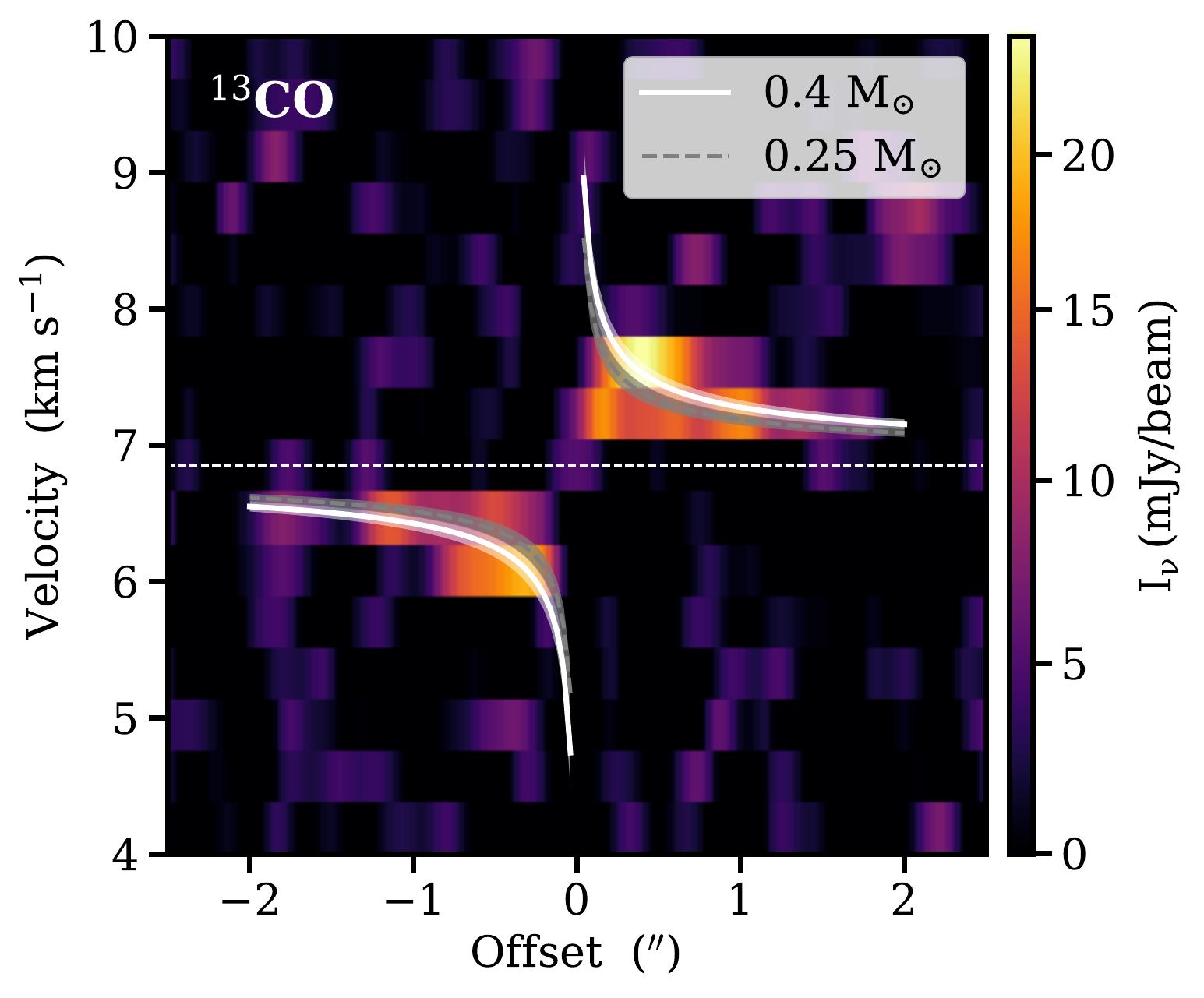} \\
    \caption{The position-velocity diagram of  $^{13}$CO emission along the disk major axis. Keplerian rotation models around a central star of 0.4 and 0.25\,$M_{\odot}$ are shown as white and grey curves, respectively. The shaded regions indicate the effect of disk inclination angle ($\pm2\degr$). \label{fig:co_pv}}
\end{figure}

As the $^{13}$CO is less affected by nearby clouds, we use its velocity field to roughly assess the mass of the central star. Figure~\ref{fig:co_pv} shows the position-velocity diagram of the $^{13}$CO emission, derived along the disk major axis (PA of 122.7$\degr$) with a \texttt{width=`1.5arcsec'} in \texttt{CASA} task \texttt{impv}. With the lack of emission at high-velocity channels, it is difficult to determine the stellar dynamical mass in high precision. Based on current observations, any value from 0.25--0.4\,$M_{\odot}$ is reasonable.

%%%%%%%%%%%%%%%%%%%%%%%%%%%%%%%%%
%%%%%%%%%%%%%%%%%%%%%%%%%%%%%%%%%

%%%%%%%%%%%%%%%%%%%%%%%%%%%%%%%%%

\bibliography{sample631}
\end{CJK*}
\end{document}